\documentclass[iop]{emulateApJ}
\usepackage{graphicx}
\usepackage{txfonts}
\usepackage{amssymb}
\usepackage{amsfonts}
\usepackage{color}
\usepackage{multirow}
\usepackage[normalem]{ulem}
\usepackage{natbib}
\usepackage{rotating}

\newcommand\T{\rule{0pt}{2.6ex}}

\newcommand\B{\rule[-1.2ex]{0pt}{0pt}}

\shorttitle{The high-energy emission of GX 339--4}

\begin{document}


\title{Variability and spectral modeling of the hard X-ray emission of GX 339--4 in a bright low/hard state}


\author{R. Droulans\altaffilmark{1}, R. Belmont\altaffilmark{1}, J. Malzac\altaffilmark{1} and E. Jourdain\altaffilmark{1}}
\affil{Universit\'e de Toulouse ; UPS ; CESR ; 9 avenue du Colonel Roche, F-31028 Toulouse, France \\
CNRS ; UMR5187 ; F-31028 Toulouse Cedex 9, France}

\begin{abstract}
We study the high-energy emission of the Galactic black hole candidate
GX 339--4 using \textit{INTEGRAL}/SPI and simultaneous \textit{RXTE}/PCA data.
By the end of January 2007, when it reached its peak 
luminosity in hard X-rays, the source was in a bright hard state. The SPI data from this period
show a good signal to noise ratio, allowing a detailed study of the spectral energy distribution up to several hundred keV.
As a main result, we report on the detection of a variable hard spectral feature ($\geq$$150$\,keV) which represents a significant excess with respect to the cutoff power law shape of the spectrum. The SPI data suggest that the intensity of this feature is positively correlated with the $25$\thinspace --\thinspace $50$\,keV luminosity of the source and the associated variability time scale is shorter than $7$\,hours.
The simultaneous PCA data, however, show no significant change in the spectral shape, indicating that the source is not undergoing a canonical state transition.
We analyzed the broad band spectra in the lights of several physical models, assuming different heating mechanisms and properties of the Comptonizing plasma.
For the first time, we performed quantitative model fitting with the new versatile Comptonization code \textsc{belm}
, accounting self-consistently for the presence of a magnetic field. We show that a magnetized medium subject to pure non-thermal electron acceleration provides a framework for a physically consistent interpretation of the observed $4$\thinspace --\thinspace $500$\,keV emission. Moreover, we find that the spectral variability might be triggered by the variations of only one physical parameter, namely the magnetic field strength.
Therefore, it appears that the magnetic field is likely to be a key parameter in the production of the Comptonized hard X-ray emission.
\end{abstract}


\section{Introduction}

GX 339--4 was discovered in the early 70's by the MIT X-ray detector aboard the OSO-7 mission \citep{markert1973}.
The source is classified as a low mass X-ray binary (LMXB; \citealt{shahbaz2001}) from upper limits on the optical luminosity of the 
companion star. It is believed to harbor
a black hole, for which \citet{hynes2003} derived a mass function of $5.8 \pm 0.5\,\rm M_{\sun}$.
The inclination of the system is yet uncertain. A study of the binary parameters \citep{zdziarski2004} revealed a plausible lower limit of $i \geq 45^{\circ}$, while 
\citet{cowley2002} suggested $i \leq 60^{\circ}$ because the system is not 
eclipsing.
On the other hand, spectral fits of the Fe $K_{\alpha}$ region with Chandra \citep{miller2004a} and XMM Newton \citep{miller2004b,reis2008}
clearly favor lower values for the inner disc inclination (typically $i \sim 20^{\circ}$).
This may indicate that the 
inner accretion disc is warped, making it appear at a lower inclination than the orbital plane. Similarly, there is no certainty concerning the distance to the source. Resolving the velocity structure along the line of sight, \citet{hynes2004} obtained a conservative lower limit of $d \geq 6$\,kpc. \citet{zdziarski2004} analyzed the binary parameters of the system and found the most plausible distance to be around $8$\,kpc. In this work, we use $d = 8$\,kpc, $i = 50^{\circ}$ and, assuming a small companion mass, we infer $M=13\,\rm M_{\sun}$ from the mass function.

Black hole binaries (BHBs) are powerful engines producing high-energy radiation up to the $\gamma$-ray domain.
From an observational perspective, they appear in four spectral states, namely, quiescent, low/hard, intermediate and high/soft (\citealt{tanaka1995}, \citealt{belloni2005}; see also \citealt{mr06} for a slightly different classification).

{\it Soft states:}
In the soft state, the energy spectrum is dominated by a soft ($\leq 10$\,keV) component which is attributed to thermal emission from an optically thick geometrically thin accretion disc \citep{ss73} extending down to the last stable orbit. 
Above $10$\,keV, the emission is characterized by a complex hard X-ray continuum with no clear presence of a high-energy cutoff (see e.g. \citet{gierlinski1999,motta2009,caballero-garcia2009}). This component can be attributed to inverse Compton scattering of soft photons (UV, soft X) in a hybrid thermal/non-thermal electron plasma (the so-called 'corona') (see e.g. the review by \citealt*{dgk07}, hereafter DGK07).

The non-thermal particles in the corona may be generated due to the magnetic field. Indeed, the Parker instability is able to transport a significant fraction of the accretion power above and below the disc \citep{galeev1979,uzdensky2008}, where the energy may then be dissipated through magnetic reconnection. Alternatively, Fermi acceleration at relativistic shocks is also expected to produce non-thermal particle distributions which can explain the observed steep power law emission in hard X-rays.

{\it Hard states:}
In the hard state, the energy spectrum is very different. The disc emission almost vanishes and the spectrum is dominated by a hard power law component (photon index $\Gamma =$ 1.4 -- 2.0) with a nearly exponential cutoff ($E_{\rm cut}$ = 50 -- 150\,keV) (see e.g. \citealt{zdziarski1998}). Such a spectrum is well described by thermal Comptonization in a hot, optically thin electron-proton plasma \citep{sunyaev1979,zg04}. 
Soft $\gamma$-ray observations of several BHBs additionally revealed a high-energy excess with respect to a thermal Comtonization model, suggesting that in some cases at least some level of non-thermal electron acceleration is also required. For instance, COMPTEL observed a non-thermal tail in the averaged hard state spectrum of Cygnus X-1 \citep{mcconnell2002} while OSSE and SPI detected such a feature during bright hard states of GX 339-4 \citep{johnson1993,wardzinski2002,joinet2007}.

{\it Intermediate states:}
Intermediate states are observed during transitions between the hard and the soft states
and usually show the characteristic features of both \citep{miyamoto1991,mendez1997,kong2002}. For GX 339--4, \citet{belloni2005} defined two different varieties of intermediate state: the hard intermediate state (HIMS) and the soft intermediate state (SIMS). The transition from the HIMS to the SIMS of the 2004 outburst of GX 339--4 is reported in \citet{delsanto2008}.

According to a popular scenario, the different spectral states can be explained through changes in the geometry of the accretion flow.
The weakness of the thermal component in the hard state is generally interpreted as a consequence of a truncated accretion disc (DGK07), which, in its inner parts, is replaced by a hot, advection dominated accretion flow (ADAF; \citealt{shapiro1976,narayan1994,yuan2007}). 
In these solutions, gravitational energy is converted into thermal energy of protons, which in turn heat the electrons through Coulomb collisions. This process naturally forms the quasi thermal electron distributions that are required to explain the typical hard state spectra. Moreover, recent models of hot accretion flows include a small non-thermal component that can account for the sometimes observed high-energy excess.

However, a number of recent results seem to question this paradigm. First, it is notoriously difficult to estimate the inner disc radius in the hard state and its recession is a highly debated topic \citep[see e.g.][]{cabanac2009}. Although several reports support the truncated disc scenario \citep{tomsick2009,done2009}, others suggest that even in hard states the accretion disc may extend down to a few Schwarzschild radii \citep{miller2006,reis2008}, which would be inconsistent with the standard ADAF scenario. Second, hot accretion flows generally exist only if the optical depth is small ($\tau_T \ll1$), which implies that the electron-proton coupling is weak and therefore allows high proton temperatures. However, it is difficult to align such a small optical depth with both the hard X-ray spectral slope and the thermal cutoff energy. When the spectral slope is reproduced, classic ADAF models lead to higher electron temperatures than those observed in BHBs, as it was shown for Cygnus X-1 or GX 339-4 \citep[hereafter MB09]{yuan2004,mb09}. Recent Monte-Carlo simulations, which account for global Compton cooling and general relativistic effects \citep{xie2010}, yield self-consistent solutions in which the peak energy of the spectrum is reduced with respect to previous results. However, it is still not straightforward to accommodate the inferred high-energy cutoff to the observations, even if gravitational redshift is accounted for.

As an alternative to the ADAF-like models, the Comptonizing medium in the hard state could as well be powered by the same non-thermal mechanisms that are believed to accelerate the electrons in the soft state, i.e. diffusive shock acceleration or magnetic reconnection. Such a model naturally accounts for the presence of a non-thermal component in the high-energy spectrum.
In addition, MB09 showed that the steady state electron distribution can appear quasi thermal even if acceleration mechanisms are purely non-thermal (see also \citet{poutanen2009}). These authors studied the thermalizing effects of the magnetic field, since it was pointed out by \citet*{ghisellini1998} that the very fast emission and absorption of synchrotron photons (the so-called 'synchrotron boiler effect') is able to thermalize the electron distribution in a few light-crossing times. Using \textsc{belm}, a new code which includes this effect \citep*{bmm08}, MB09 qualitatively explained the variety of spectral states observed in the prototypical BHB Cygnus X-1. The model is consistent with a disc recession in the hard state \citep{poutanen2009}, but since the weakness of the soft component could as well result from a lower disc temperature \citep{beloborodov1999,malzac2001}, it does not necessarily require a change in the geometry of the accretion flow (MB09).

In this paper, we use observations of GX 339--4 with SPI/\textit{INTEGRAL} and PCA/\textit{RXTE} to study the spectral energy distribution of the bright hard state from the 2007 outburst. We analyze the spectral variability in the framework of the models cited herebefore.  
The data and the associated reduction procedures are described in Section 2. Section 3 is dedicated to a phenomenological analysis of the high-energy behavior while in Section 4 we present an extensive physical analysis of the broad band spectra. The results and their implications are discussed in Section 5 and summarized in the conclusive section of the paper.

\section{Observations}

\begin{figure}
  \centering
  \includegraphics[width=\columnwidth]{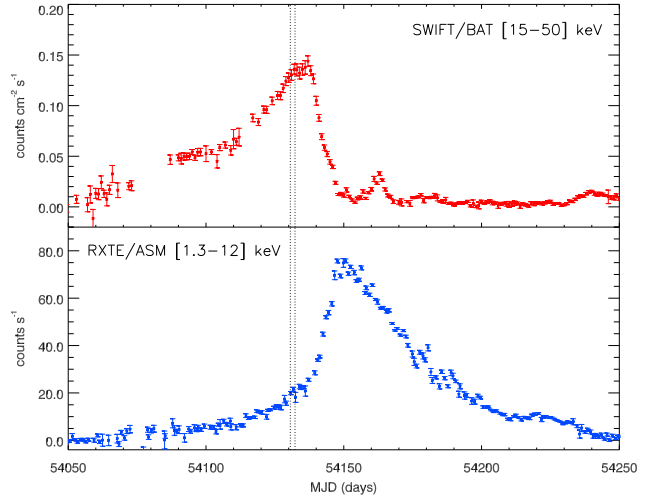}
     \caption{Overview of the 2007 outburst showing the daily light curves of GX 339--4 obtained by \textit{SWIFT}/\textit{BAT} 
	(15 -- 50\,keV ; top panel) and \textit{RXTE}/\textit{ASM} (1.3 -- 12\,keV ; lower panel). The vertical dotted lines indicate the time period during which the here presented SPI data were obtained.}
     \label{outburst}
\end{figure}

The here presented data on GX 339--4 were obtained with the \textit{INTEGRAL} and \textit{RXTE} observatories
at the end of January 2007. At that time, the source was very bright in hard X-rays (cf. Figure \ref{outburst}) and \citet{motta2009} identified its spectro-temporal properties to be characteristic of the hard state.

\subsection{Instruments}

Our analysis focuses on the results obtained by the spectrometer SPI \citep{vedrenne2003}, which is one of the two main instruments
aboard the international gamma ray observatory \textit{INTEGRAL}. Operating in the 20\,keV -- 8\,MeV band, 
SPI uses germanium detectors to provide high spectral resolution while imaging is performed through the coded mask technique (Roques et al. 2003).
The observational strategy of the \textit{INTEGRAL} mission is to sample the region of interest by means of 30 -- 40 minute long fixed pointings (the so-called science-windows), each separated by a $2^{\circ}$ angular distance (see \citet{jensen2003} for details).
In order to extend the spectral coverage to lower energies, we also considered 
simultaneous 4 -- 28\,keV data from the Proportional Counter Array (PCA) on \textit{RXTE} \citep{bradt1993}.
\begin{table}
\centering
\begin{tabular}{ccccc}
\hline\hline
\T \B Instruments & Obs. ID & MJD start & MJD stop & Exp time (ks)
\\
\hline
\T SPI/IBIS & 525 & $54130.6$ & $54132.2$ & $107$ \\
SPI/IBIS & low (cutoff) & --- & --- & $36$ \\
SPI/IBIS & high (excess) & --- & --- & $36$ \\
PCA/HEXTE & 92035-01-01-02 & $54131.10$ & $54131.17$ & $3.7$ \\
PCA/HEXTE & 92035-01-01-04 & $54132.08$ & $54132.15$ & $3.7$ \\
\hline
\label{tab1}
\end{tabular}
\caption{\rm{Log of the different observations discussed in the paper.}}
\end{table}

\subsection{Data and reduction methods}

The SPI data from the GX 339--4 region were obtained during \textit{INTEGRAL} revolution 525, 
lasting from 2007 January 30 to February 1 (cf. Table 1).
We selected the science-windows where GX 339--4 was less than $12^{\circ}$ off 
the central axis and which did not show any contamination by solar flares or
radiation belt exit/entry. This resulted in a set of 50 useful pointings representing $107$\,ks of observational coverage. 
In order to determine the emitting sources in the field of view,
we performed 25 -- 50\,keV 
imaging with the \textsc{spiros} software \citep{skinner2003}.
Aside from the main source, 4U 1700--377, OAO 1657--415, IGR J16318--4848 and GX 340+0 were detected above a $5\,\sigma$ threshold. The positions of these sources were then given as a priori information to a specific flux-extraction algorithm, using the SPI instrument response for sky-model fitting. 
In a first run of the software, we allowed the background
normalization as well as the source fluxes to vary between successive pointings.
In this way we determined the most appropriate variability
time scale for each component. The background 
normalization was more or less stable during the SPI observation,
therefore we assumed no background variability in the final reduction process.
Some of the sources in the field, however, showed considerable flux variations. Most notably, 4U 1700--377 was extremely variable, with a
20 -- 50\,keV flux per science-window ranging from $0$ to $700$\,mCrab.

There are two \textit{RXTE} observations which are simultaneous with the \textit{INTEGRAL} exposures (see Table 1). The first observation took place towards the middle and the second towards
the end of revolution 525 (cf. the shaded areas in Figure \ref{lc}).
For both observations, we downloaded the standard products from the HEASARC website\footnote{http://heasarc.gsfc.nasa.gov/docs/archive.html} and analyzed the data in \textsc{xspec} v11.3.2 \citep{arnaud1996}.
Apart from a global $3$ per cent flux difference, the PCA spectra from both observations are compatible within the error bars and were
co-added using the \textsc{addspec} routine from the \textsc{ftools} package. Following \citet{motta2009}, we added a systematic error of $0.6$ per cent to each PCA channel.

\subsection{SPI light-curve and data subsets}

\begin{figure}
  \centering
  \includegraphics[width=\columnwidth]{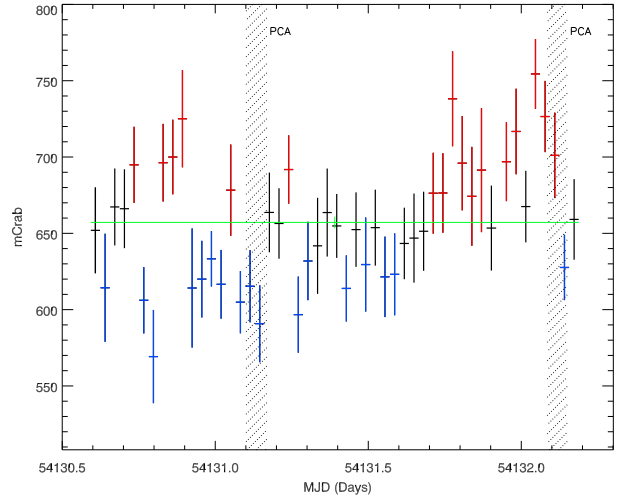}
     \caption{25 -- 50\,keV SPI light curve from revolution 525. The pointings selected for the high and low flux data sets are shown in red and blue, respectively. The cyan line represents the average flux from the SPI observation. The shaded areas indicate the time periods of the simultaneous PCA observations.}
     \label{lc}
\end{figure}

Figure \ref{lc} shows the 25 -- 50\,keV SPI light curve from GX 339--4 obtained during revolution 525. Each bin represents 
one pointing or science-window, which corresponds to a time scale of about $40$ min.
Overall, the 25 -- 50\,keV flux shows no particular evolution, 
indicating that the source remained in the bright hard state during the $\sim$$1.5$ day long observation. 
However, the light-curve reveals minor variability (on a time scale of hours, or less) around the average flux value 
($<$$F$$>=657.1 \pm 3.8$\,mCrab). To investigate
whether this variability could be linked with changes in the spectral energy distribution,
we fixed two arbitrary flux bounds ($F_{\rm low}=640$\,mCrab and $F_{\rm high}=670$\,mCrab) and grouped
the science-windows where $F_{\rm scw}<F_{\rm low}$ and $F_{\rm scw}>F_{\rm high}$, respectively.
For each of the 
two resulting data subsets, color-coded in blue and red and referred to as \textit{low} and \textit{high} respectively, we then produced the averaged 25 -- 500\,keV spectrum. Both subsets consist of an equal number of science-windows, which allows a straightforward comparison of the spectra.
Since the source flux is outstandingly high, the spectral extraction yields significant results, even if the number of science-windows is relatively small.

\section{High energy emission}

\subsection{SPI spectra}

The resulting 25 -- 500\,keV spectra differ by $\sim$$14$ per cent in terms of their total flux and were separately fitted in \textsc{xspec} with a simple cutoff power law model. The uncertainty on the model parameters is given at the $90$ per cent confidence level ($\Delta$$\chi^2=2.7$).
As the SPI data alone do not allow to simultaneously constrain the photon index and the cutoff 
energy, we fixed the former to $\Gamma=1.3$ and left the latter free to vary.
The model provides a good fit to the \textit{low} spectrum ($\chi^2/22 =0.75$) 
and we infer a significant cutoff at $E_{\rm c}=58.6\pm2.2$\,keV.
For the \textit{high} spectrum, the marked curvature around $40$\,keV still suggests the presence of a cutoff energy, fitted at 
$E_{\rm c}=56.2\pm2.1$\,keV. With respect to the model, however, we observe a significant excess above $150\,$keV, leading to an
unacceptable quality of the fit ($\chi{{}^2}/28 =1.82$).
To account for the high-energy excess, we added a second power law of fixed photon index 
$\Gamma=1.6$\footnote{If this parameter is left free to vary, the fitted $90$ per cent confidence interval is $\Gamma \in [0.3 , 1.9]$}. The presence of this second component, which reduces the cutoff energy to $E_{\rm c}=49.5\pm3.8$\,keV but improves the fit to $\chi{{}^2}/27 =1.03$, is statistically required as shown by the \textsc{ftest} \citep{bevington1992}. Indeed, we infer a probability of $P_{\textsc{ftest}} = 6.1 \times 10^{-5}$ that the improvement of the fit was a chance event.
As a consequence, we conclude that both SPI spectra essentially differ in terms of the highest energy emission ($>150\,$keV).

\subsection{Imaging and spectral robustness}

As the flux extraction process of the SPI data can be sensitive to background modelling and a priori information about the 
distribution of emitting sources in the field of view, we double checked our results.
In the 150 -- 450\,keV \textsc{spiros} image drawn from the \textit{high} data set, GX 339--4 
is detected without any a priori information at a flux level of $347.4 \pm 49.9$\,mCrab and a $7.0\, \sigma$ significance. 
This contrasts to a non-detection in the \textit{low} data set, where the flux significance at the source position is below $2.5\, \sigma$.
We note that the maximum level of the uniformly distributed residuals is around $ 4\, \sigma$ in both images.

Since GX 339--4 is the only source detected above $150$\,keV,
we used a sky model with a single point source and re-extracted the high-energy part ($>$$150\,$keV) of the spectra.
We also tested different background models along with various variability time scales.
The obtained results are all perfectly compatible within the $1\, \sigma$ errors, showing that the spectra are insensitive to the details of the flux extraction process.
We conclude that the respective presence of a cutoff and a high-energy excess in the SPI data is robust and significant.
As a consequence, we rename the \textit{low} and \textit{high} spectra to \textit{cutoff} and \textit{excess}, respectively.
\begin{figure}
  \centering
  \includegraphics[width=\columnwidth]{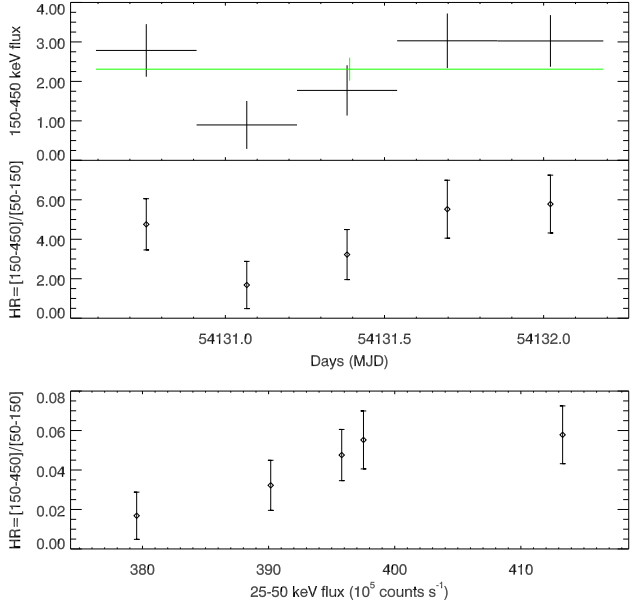}
  \caption{\textit{Top panel}: 150--450\,keV SPI lightcurve with a binning of $7$\,h. The flux is given in units of $10^5$ counts s$^{-1}$. The cyan cross indicates the average flux and flux uncertainty. \textit{Middle panel}: [150--450]/[50--150]\,keV hardness ratio as a function of time. \textit{Lower panel}: [150--450]/[50--150]\,keV hardness ratio as a function of the 25 -- 50\,keV flux.}
     \label{hardness}
\end{figure} 

\subsection{Time evolution}

The presence and absence of a high-energy excess in the high and low flux spectra suggests that the appearance of this feature is correlated with the 25 -- 50\,keV flux of the source. To inverstigate this issue further, we extracted light curves in the 25 -- 50, 50 -- 150 and 150 -- 450\,keV energy bands with a binning of $10$ science-windows ($\approx$$7$\,h) and plotted the [150 -- 450]/[50 -- 150]\,keV hardness ratio as a function of time and as a function of the 25 -- 50\,keV flux. From these plots, shown in Figure \ref{hardness}, it is clear that the strength of the high-energy excess is correlated with the 25 -- 50\,keV flux. Since a different time binning ($8$ or $12$ science-windows for instance) leads to the same results and since the 25 -- 50\,keV flux is a good tracer of the total X-ray luminosity, we conclude that the intensity of the high-energy tail is most likely correlated to the total X-ray luminosity of the source.

For statistical resons, the SPI data do not allow to quantify the exact time scale of the phenomenon. Nevertheless, from the 25 -- 50\,keV science-window light curve (cf. Figure \ref{lc}) and the above analysis, we can assess that the time scale of the observed spectral variability is shorter than $7$\,h.

\subsection{Cross-check with other instruments}
\begin{figure*}
\begin{minipage}{\textwidth}
\begin{tabular}{cc}
  \includegraphics[width=8.55cm]{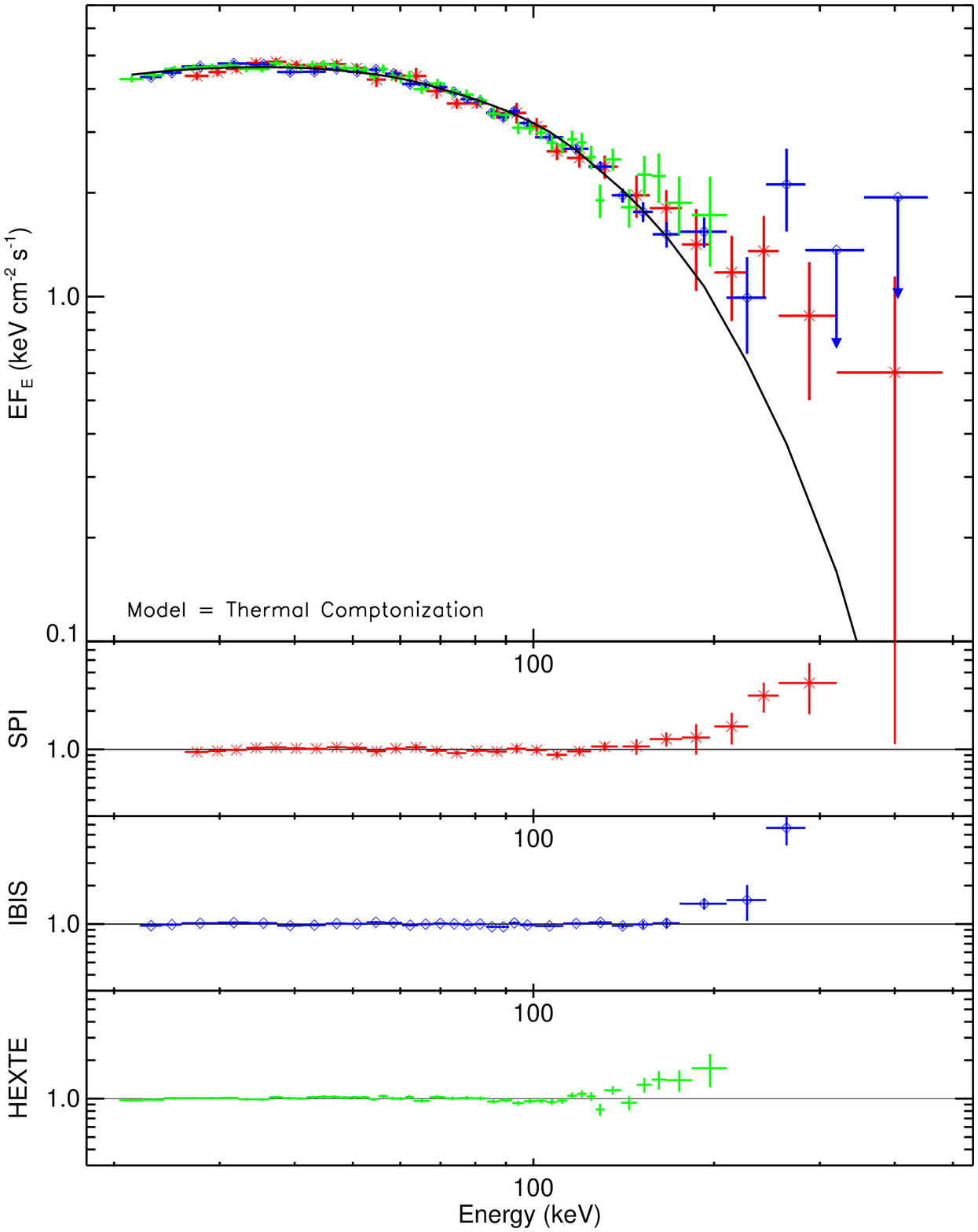} &
  \includegraphics[width=8.55cm]{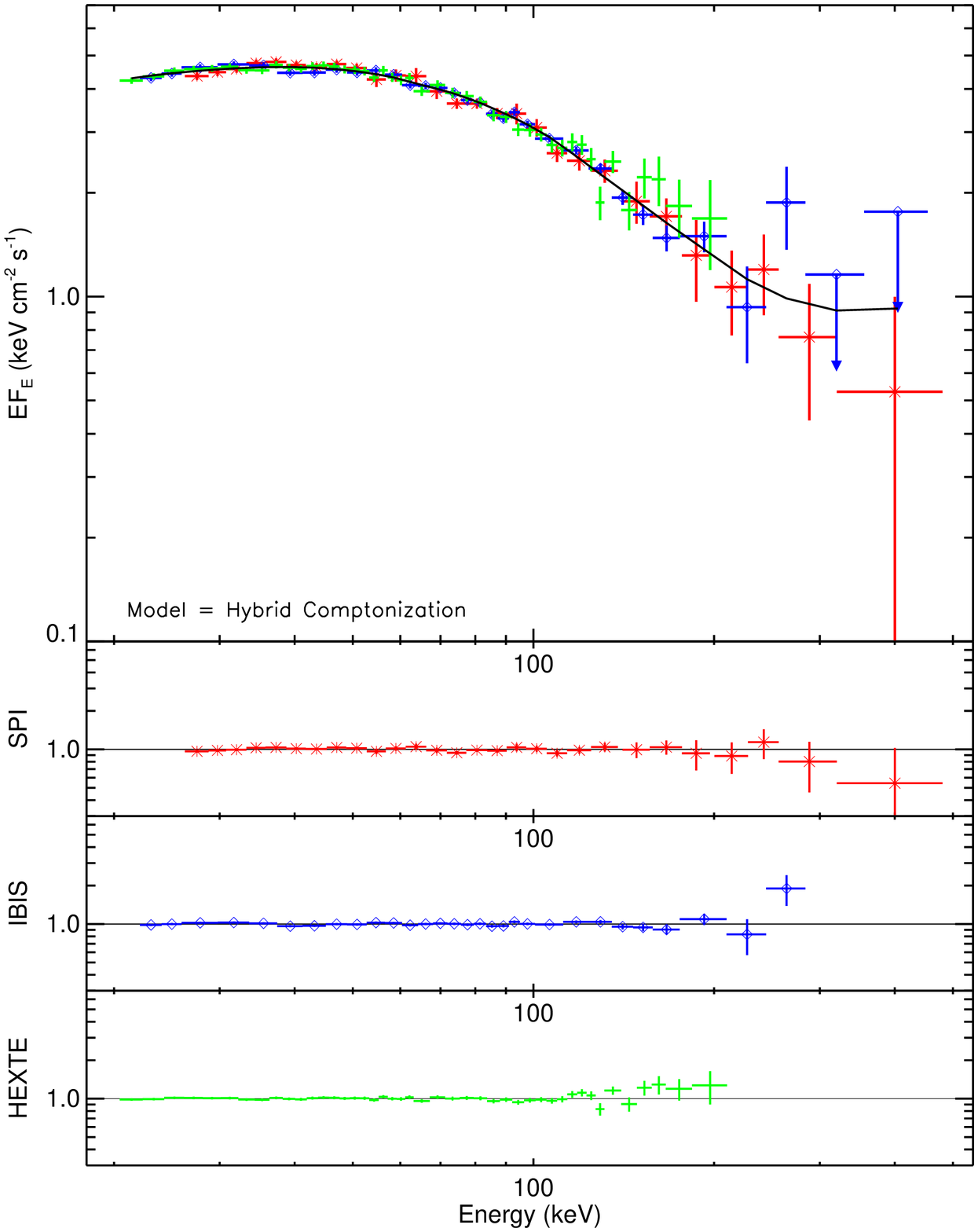}
\end{tabular}
  \centering
  \caption{Total averaged SPI (red), IBIS/ISGRI (blue) and HEXTE (green) spectra, jointly fitted with a thermal (left) and a hybrid thermal/non-thermal (right) Comptonization model. The normalization was fixed to the calibration of SPI and we applied mutiplicative constants of $C_{\rm ISGRI}=1.03$ and $C_{\rm HEXTE}=1.10$ to the IBIS/ISGRI and HEXTE spectra, respectively. The lower panels indicate the data/model ratios for each instrument separately. These show that a pure thermal Comptonization model is not able to account for the emission above $150$\,keV.}
\label{total}
\end{minipage}
\end{figure*}

\begin{table}[h!]
\centering
\begin{tabular}{ccccc}
\hline\hline
 \T \B Instrument & Model & $(\chi^2/\nu)_{\rm i}$ & $(\chi^2/\nu)_{\rm f}$ & P$_{\textsc{ftest}}$  \\
\hline
\T \multirow{2}*{SPI} & \textsc{cutoffpl} & $36/26$ & $18/25$ & $4 \times 10^{-5}$ \\
 & \textsc{eqpair} & $47/26$ & $22/25$ & $2 \times 10^{-5}$ \\
\T \multirow{2}*{IBIS/ISGRI} & \textsc{cutoffpl} & $32/29$ & $26/28$ & $2 \times 10^{-2}$ \\
 & \textsc{eqpair} & $39/29$& $25/28$ & $5 \times 10^{-4}$ \\
\T \multirow{2}*{HEXTE} & \textsc{cutoffpl} & $59/44$ & $47/43$ & $2 \times 10^{-3}$ \\  
& \textsc{eqpair}& $66/44$ & $47/43$ & $1 \times 10^{-4}$ \\
\T \multirow{2}*{SPI+IBIS+HEXTE} & \textsc{cutoffpl} & $145/103$ & $105/102$ & $1 \times 10^{-8}$ \\  
& \textsc{eqpair}& $141/103$ & $99/102$ & $2 \times 10^{-9}$ \\
\hline
\label{ftests}
\end{tabular}
\caption{Quantification of the high-energy excess in the total averaged SPI, IBIS/ISGRI and HEXTE spectra with respect to a cutoff power law and a thermal Comptonization model. To account for the high-energy excess, we added a second power law component with a fixed spectral index of $\Gamma=2.0$ in the phenomenological model, while we allowed for non-thermal heating in the physical model. Each time we give the fit quality before $(\chi^2/\nu)_{\rm i}$ and after $(\chi^2/\nu)_{\rm f}$ adding a degree of freedom, along with the \textsc{ftest} probability that the fit improvement was a chance event.}
\end{table}
Since the detection of a high-energy tail is a critical issue, we cross-checked the SPI results with other instruments. We analyzed the data obtained simultaneously by the soft gamma ray imager IBIS/ISGRI \citep{ubertini2003} aboard \textit{INTEGRAL} and the high energy X-ray timing experiment HEXTE aboard \textit{RXTE}. The HEXTE spectra from the two simultaneous \textit{RXTE} observations (cf. Table 1) have been presented by \citet{motta2009}, who kindly provided us the reduced data. Both spectra are compatible within the $1\sigma$ errors and were co-added to obtain better statistics at high energies. 
The total SPI and IBIS/ISGRI spectra (averaged over the whole \textit{INTEGRAL} revolution 525) have been presented by \citet{caballero-garcia2009}, who report on the detection of a high-energy excess above 150\,keV. We re-extracted the IBIS/ISGRI data using the standard OSA 8.0 software package\footnote{available from the \textit{INTEGRAL} Science Data Centre (ISDC)} and jointly fitted the averaged SPI, IBIS/ISGRI and HEXTE spectra in \textsc{xspec}. We find a good agreement between the three instruments, not only in spectral shape but as well in normalization. In addition, the data from all three instruments confirm the presence of a high-energy excess with respect to a phenomenological cutoff power law and a thermal Comptonization model (cf. Figure \ref{total}, Table 2). For the latter we used \textsc{eqpair} \citep[cf. section 4.2]{coppi1999} with non-thermal heating turned off ($l_{\rm nth}=0$). The reflection amplitude is poorly constrained and was therefore fixed to $\Omega/2\pi=0.35$ (average value derived from the physical analysis, cf. Table 3).

The two IBIS/ISGRI spectra averaged separately over the \textit{low} and \textit{high} data sets are consistent with the SPI results. However, due to the shorter exposure times, the high-energy excess is no longer required in the ISGRI data. This is not surprising since the sensitivity of ISGRI above 200\,keV is lower than the sensitivity of SPI. 
As a consequence, we did not consider the IBIS/ISGRI data in the remainder of the paper because they do not improve the high-energy constraints with respect to the models.
The HEXTE data, on the other hand, are obtained with even shorter exposure times and are thus unsuitable for our group-wise spectral analysis.

\section{Broadband spectral analysis}

In order to understand the processes that are likely to cause
the observed X/$\gamma$-ray emission, we analyzed the broad band PCA/SPI spectra in the lights of different physical Comptonization models.
Since the spectral variability is occurring at high energies only, each one of the two SPI spectra (cf. section 3.1) was jointly fitted with the co-added PCA spectrum. A variable multiplicative factor was added to account for cross-calibration uncertainties as well as the luminosity difference between the two SPI spectra. The difference to unity of this factor is never larger than $8$ per cent. Except for slight changes in the $\chi^{2}$ values, neither the qualitative nor the quantitative results are affected when using one of the single PCA spectra instead of the co-added spectrum. 

\subsection{General model}

The high-energy source in GX 339--4 is modeled by a spherical, magnetized, fully ionized proton-electron/positron plasma of radius $R$. 
The emission of the plasma is derived by self-consistent computations of the equilibrium electron distribution accounting for Compton scattering, synchrotron emission/absorption, pair production/annihilation and Coulomb collisions (\textit{e-e} and \textit{e-p}). The Thomson optical depth of the plasma is given by $\tau_{\rm T}=\tau_{\rm ion}+\tau_{\rm pair}$, where $\tau_{\rm ion}$ represents the optical depth of ionization electrons and $\tau_{\rm pair}$ is the opacity that arises from pair production.

The plasma properties essentially depend on the magnetic field strength as well as on the power supplied by external sources.
The tangled magnetic field strength is parameterized by the magnetic compactness:
\begin{equation}
l_{B}=\frac{\sigma_{\rm T}}{m_{\rm e}c^2}R\frac{B^2}{8\pi}.
\end{equation}
Energy injection is quantified by the compactness parameter:
\begin{equation}
l=\frac{\sigma_{\rm T}}{m_{\rm e}c^3}\frac{L}{R},
\end{equation}
where $L$ is the power supplied to the plasma, $\sigma_{\rm T}$ the Thomson cross-section and $m_{\rm e}$ the electron rest mass. The general model comprises three possible channels for providing energy to the coupled electron-photon system: (i) non-thermal electron acceleration ($l_{\rm nth}$), (ii) thermal heating of the electron distribution ($l_{\rm th}$) and (iii) external soft radiation from a geometrically thin accretion disc ($l_{\rm s}$). The non-thermal acceleration processes are mimicked by continuous electron injection at a power law rate with index $\Gamma_{\rm{inj}}$ (i.e. $n(\gamma) \propto \gamma^{-\Gamma_{\rm inj}}$) and Lorentz factors ranging from $\gamma_{\rm{min}}$ to $\gamma_{\rm max}$. The thermal heating of the electron distribution could be caused by Coulomb interactions with a population of hot ions.
The incident radiation from an accretion disc is modeled by a blackbody component of temperature $kT_{\rm bb}$. 
Since the temperature of this component cannot be constrained by our data, we follow \citet{delsanto2008} and fix $kT_{\rm bb}$ to the fiducial value of $300$\,eV. 
All the injected energy ends up into radiation, and in the compactness formalism the total radiated power in steady state holds: $l=l_{\rm nth}+l_{\rm th}+l_{\rm s}$.

For a given source flux $F$, the total compactness can be estimated by the formula:
\begin{equation}
l = 100 \left(\frac{F}{F_0}\right) \left(\frac{d}{8 \rm{kpc}}\right)^2 \left(\frac{13 M_{\sun}}{M}\right) \left(\frac{30R_{\rm G}}{R}\right)
\label{comp}
\end{equation}
where $M$ is the mass of the black hole and $d$ the distance to the source. 
Here we expressed $l$ in terms of $F_{0}=2.6 \times 10^{-8}$\,erg s$^{-1}$cm$^{-2}$, the observed average 4 -- 500\,keV flux of GX 339--4.
Assuming a spherical plasma of radius $R=30$\,R$_{\rm G}$, a black hole mass of $M=13$\,M$_{\sun}$ and a distance of $d=8$\,kpc,
this leads to a total compactness of $l \sim 100$.

Besides being a source of soft seed photons, the accretion disc may also give rise to Compton reflection of the incident hard X-rays from the Comptonizing plasma. 
The reflected emission is calculated using the viewing-angle
dependent Green's functions \citep{magdziarz1995}, but accounting for general relativistic effects. The spatial extension of the possibly ionized disc is parameterized by its inner and outer radii, fixed at $R_{\rm in}=6$\,R$_{\rm G}$ and $R_{\rm out}=400$\,R$_{\rm G}$. 

Moreover, the model takes into account the K$\alpha$ fluorescence of the Fe elements in the disc (\textsc{diskline}, \citealt{fabian1989}) as well as the interstellar absorption (\textsc{phabs}). For the latter, we follow \citet{reis2008} and assume a fixed neutral hydrogen column density of
$N_{\rm H} = 5.2 \times 10^{21}\, \rm{cm}^{-2}$.

Since our data sets are energy and sensitivity limited and since such a complex modelling is partly degenerate, spectral fitting is unable to provide simultaneous constraints to all the parameters. Therefore, starting from the general model, we define several canonical sub-models where some of the parameters are held invariant.

\subsection{Thermal heating}

Given that the relative ratio of the heating mechanisms (thermal versus non-thermal) is very difficult to constrain, we investigated the extreme situations in which there is only one possible channel for providing energy to the electrons. 

First, we consider the case where the energy injection is purely thermal ($l_{\rm th}\neq 0$, $l_{\rm nth}=0$). The seed photons may arise either from an external source (i.e. the accretion disc), or, in presence of a magnetic field, from internally generated synchrotron emission. However, since the magnetic field is believed to give rise to non-thermal electron acceleration, we limited our analysis to the standard case where magnetic processes are neglected ($l_{B}=0$). 

The main model parameters are the power supplied to the Comptonizing electrons $l_{\rm th}$, the power contained in the soft photons from the disc $l_{\rm s}$, the Thomson optical depth of the plasma $\tau_{\rm ion}$, the reflection amplitude $\Omega/2\pi$, the ionization parameter of the reflecting material $\xi$ and the normalization factor between the two instruments. To calculate the model spectra, we use the hybrid thermal/non-thermal Comptonization code \textsc{eqpair} \citep{coppi1999}. A detailed description of the code and its application to Cygnus X-1 can be found in \citet{gierlinski1999}.

Since the model spectrum strongly depends on the ratio $l_{\rm th}/l_{\rm s}$, but only weakly on the absolute values of the compactness parameters, we fix $l_{\rm s}$ to a constant value in order to improve the fitting strategy. For $l_{\rm s}$ in the range $0.5<l_{\rm s}<100$, the spectral shape allows to constrain $l_{\rm th}/l_{\rm s}$ to be always of the order of $5$. Therefore we set $l_{\rm s}=15$, so that the total compactness parameter $l=l_{\rm th}+l_{\rm s}$ is consistent with Eq. \ref{comp}.

For the \textit{cutoff} spectrum, the pure thermal model delivers a good description of the data.
The best fit ($\chi^2/69 = 0.83$) yields a compactness ratio of $l_{\rm th}/l_{\rm s}=3.89^{+0.05}_{-0.08}$, a moderate opacity of $\tau_{\rm ion}=2.40^{+0.04}_{-0.08}$ and a significant ionized reflection component ($\Omega/2\pi=0.44\pm0.04$ ; $\xi=330^{+70}_{-60}$). The equilibrium temperature of the electrons is found at $32$\,keV.

For the \textit{excess} spectrum, however, the thermal model is not appropriate. The best fit is indeed of poor quality, with a reduced chi-square of $\chi^2/75 = 1.40$. We note that the degradation results only from the high-energy channels, which are responsible for $\chi^2/\nu = 65/23$ of the total $\chi^2$. Obviously, the model fails to reproduce the high-energy excess, supporting the fact that the emission above $100$\,keV is linked to non-thermal processes. We conclude that a pure thermal model is not adequate to explain the observed spectral behavior.

\subsection{Non-thermal acceleration with external Comptonization}

Now we investigate the opposite case, in other words we assume that the energy injection is purely non-thermal ($l_{\rm nth}\neq0$, $l_{\rm th}=0$).
The rest of the model remains unchanged, i.e. we consider an external source of soft photons (of fixed compactness $l_{\rm s}=15$) and we neglect the magnetic field ($l_{B}=0$).
\begin{figure*}
\begin{minipage}{\textwidth}
\begin{tabular}{cc}
  \includegraphics[width=8.55cm]{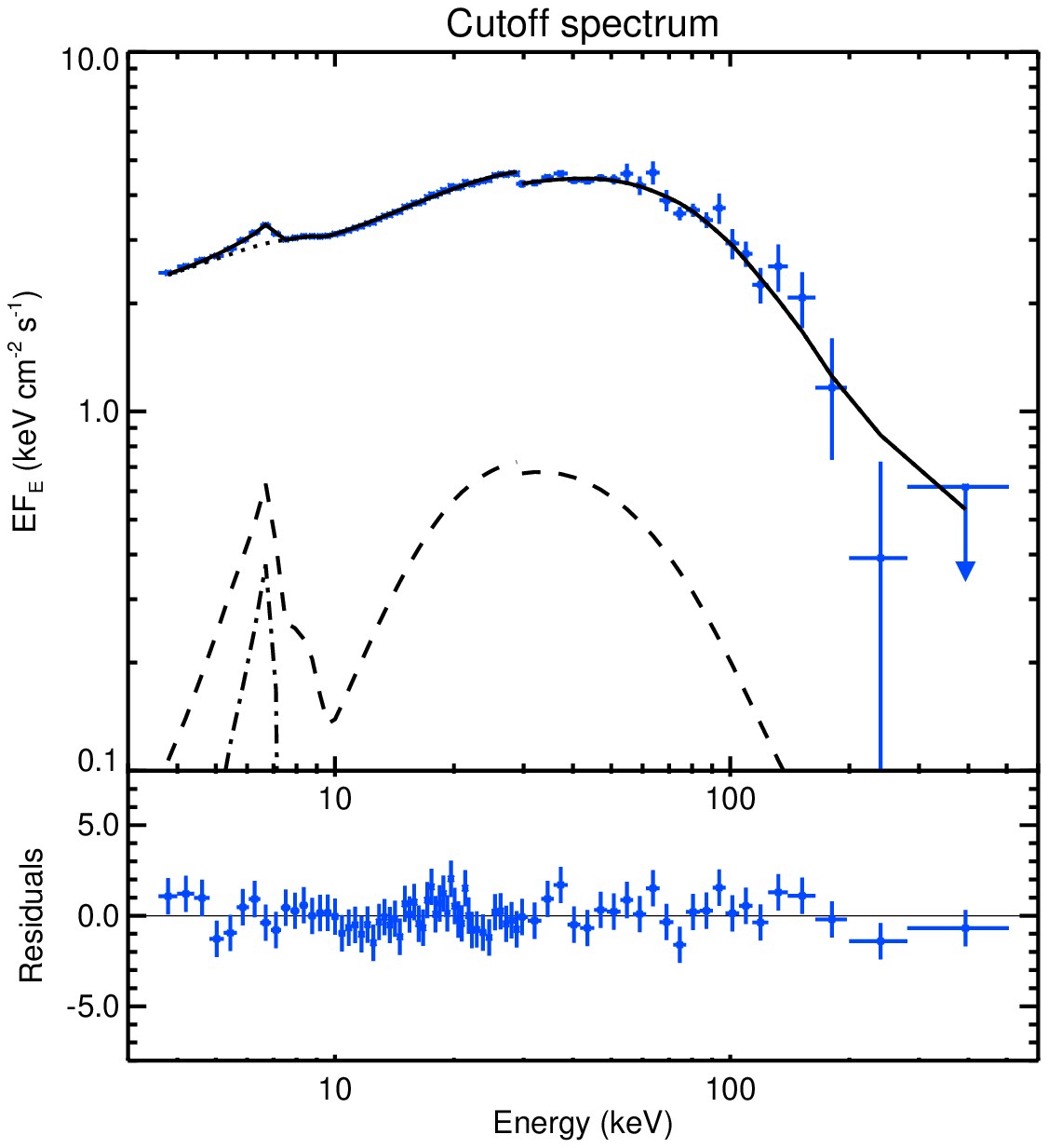} &
  \includegraphics[width=8.55cm]{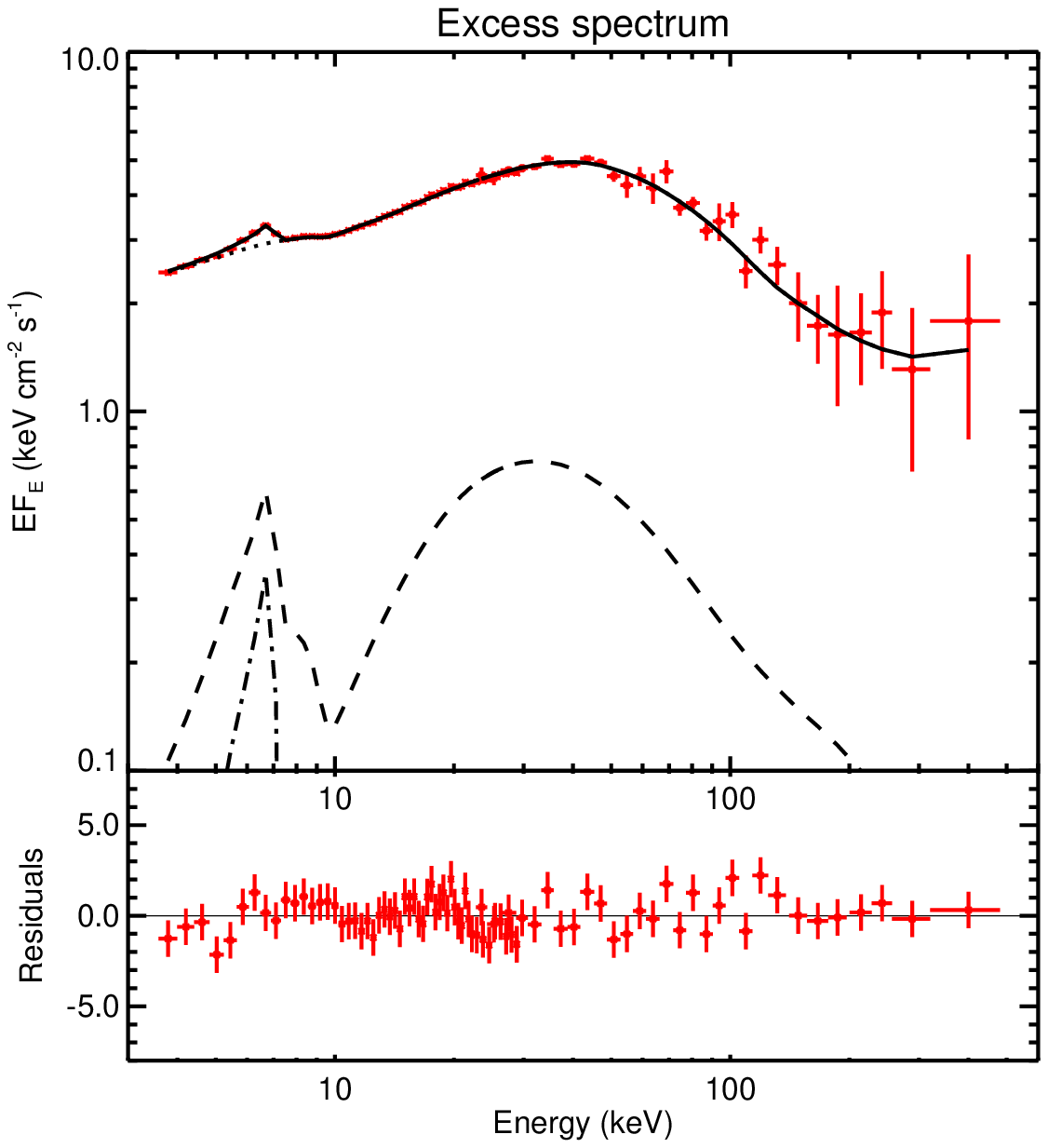}
\end{tabular}
  \centering
  \caption{Joint SPI and PCA spectra fitted with the ECM1. The total model is shown by the solid line, the dot-dashed line marks the iron fluorescence emission from the disc, the dashed line shows the reflection component and the dotted line represents the sum of the Comptonized and reflected emission. The error bars are given at the $1\sigma$ level.}\label{eq}
\end{minipage}
\end{figure*}
Since the electrons can partly thermalize through Coulomb collisions, the equilibrium distribution
is hybrid (i.e. thermal/non-thermal), even if the whole power is supplied via non-thermal injection.
The acceleration processes are phenomenologically described by the model parameters $l_{\rm nth}$, $\Gamma_{\rm{inj}}$, $\gamma_{\rm{min}}$ and $\gamma_{\rm{max}}$. Here again, we define two sub-models in oder to reduce the number of free parameters.
In the first model, we set $(\gamma_{\rm{min}},\gamma_{\rm{max}})=(1.3,1000)$, and analyze the spectra in terms of $l_{\rm nth}$ and $\Gamma_{\rm{inj}}$. This configuration is frequently used in the literature and will provide a familiar context to discuss our results.
In the second model, we adopt a more novel approach and investigate the effects of a varying maximum energy of the accelerated electrons. We keep $\gamma_{\rm min}=1.3$ but fix the spectral index to the fiducial value $\Gamma_{\rm{inj}}=2.5$, while $l_{\rm{nth}}$ and $\gamma_{\rm{max}}$ are free to vary.
For convenience, we call these models the ECM1 and the ECM2 (for External Comptonization Models), respectively. The associated model spectra are computed with the Comptonization code \textsc{eqpair} \citep{coppi1999} and the fitting results are summarized in Table 3.

\subsubsection{The injection index}

For the \textit{cutoff} spectrum, the ECM1 provides a good fit to the data ($\chi^2/68 = 0.84$, cf. Figure \ref{eq} left).
The observed cutoff shape is well reproduced by a very soft injected electron distribution. Our best fit yields a 
spectral index of $\Gamma_{\rm inj}=3.95^{+1.50}_{-0.70}$, which was fixed to this value to determine the error on the other free parameters.

We find a compactness ratio of $l_{\rm nth}/l_{\rm s}=4.35\pm0.08$, implying a non-thermal compactness of $l_{\rm nth}=65.0\pm1.0$. The optical depth of ionization electrons is fitted at $\tau_{\rm ion}=2.97\pm0.07$. Because of the soft injected electron distribution, there is only very little pair production, increasing the total optical depth to $\tau_{\rm T}=2.99\pm0.07$. The observed spectrum strongly requires a moderate amount of Compton reflection, with a fitted amplitude of $\Omega/2\pi=0.26^{+0.03}_{-0.01}$ and an ionization factor of $\xi=710\pm120$. Freezing $\Omega/2\pi$ to zero leads to a dramatically worse fit (F-test probability $<10^{-20}$).

For the \textit{excess} spectrum, the best fit ($\chi^2/74 = 1.10$, cf. Figure \ref{eq} right) requires a substantially harder power law injection, with a fitted index of $\Gamma_{\rm inj}=2.55^{+0.35}_{-0.15}$.
With respect to the \textit{cutoff} spectrum, the compactness ratio increased by $\sim$$15$ per cent to $l_{\rm nth}/l_{\rm s}=5.06^{+0.10}_{-0.08}$. The non-thermal compactness now yields $l_{\rm nth}=75.9^{+1.5}_{-1.2}$ and the total optical depth increased to $\tau_{\rm T}=3.87^{+0.07}_{-0.15}$, which is mainly due to the enhanced production of pairs. The fitted reflection amplitude and disc ionization, however, remain stable ($\Omega/2\pi=0.25^{+0.02}_{-0.03}$ ; $\xi=600\pm240$).

In order to explore the dependence on compactness, we abandoned our fiducial hypothesis $l_{\rm s}=15$ and fitted the \textit{excess} spectrum with a variable soft compactness. We find $90$ per cent confidence intervals of $0.3 < l_{\rm s} < 250$ and $4.5<l_{\rm nth}/l_{\rm s}<5.9$, which confirms that the compactness ratio does not depend on the total compactness of the source. Although our analysis is unable to uniquely determine the total compactness, the upper and lower limits on the allowed range are well constrained by the spectral shape \citep{gierlinski1999}. The upper bound represents the limit at which, despite an extremely soft injected electron distribution ($\Gamma_{\rm inj}=4.50^{+0.03}_{-0.05}$), the plasma is completely pair dominated ($\tau_{\rm ion} \sim 10^{-4}$; $\tau_{\rm T} \sim 3.0$; $kT_{\rm e} \sim 11.5$\,keV). Above this threshold, the growing amount of pairs can no longer be balanced by a softer injection spectrum, leading to an inaccurate reproduction of the thermal peak in the photon spectrum (cf. paragraph 5.2).
The lower bound corresponds to the limit at which the cooling of the high-energy particles is dominated by Coulomb instead of Compton losses, leading to an underprediction of the intensity of the high-energy tail. A change from $l_{\rm s}=0.3$ to $l_{\rm s}=0.25$ implies a degradation of $\Delta$$\chi^2=+5$ at $74$ d.o.f., showing that the threshold is well established. Hence, in the framework of an external Comptonization model involving non-thermal electron acceleration up to $\gamma_{\rm max}=1000$, the total compactness of the hard X-ray source can be conservatively constrained by means of pure spectral analysis to be $2 < l < 1500$.

\subsubsection{The maximum particle energy}

Now, in the ECM2, we freeze $\Gamma_{\rm{inj}}$ to $2.5$ but allow for variations of the maximum Lorentz factor of the accelerated electrons. For the \textit{cutoff} spectrum, we obtain again a very good fit to the data ($\chi^2/68 = 0.85$).
The effects of a much harder slope are effectively balanced by a very low cutoff energy of the injected electron distribution. Indeed, we find $\gamma_{\rm max}=4.0^{+1.3}_{-0.3}$, which means that the maximum difference of the kinetic energy between the accelerated paricles is about a factor of $10$. The other parameters are not much affected, namely we find $l_{\rm nth}/l_{\rm s}=4.25^{+0.08}_{-0.10}$ and $\tau_{\rm ion} \simeq \tau_{\rm T}=3.06^{+0.09}_{-0.06}$. Similarly, the reflection parameters remain stable ($\Omega/2\pi=0.23^{+0.03}_{-0.16}$ ; $\xi=850^{+140}_{-200}$).

For the \textit{excess} spectrum, the fixed injection index is consistent with the best fit value obtained with the ECM1. Nevertheless, it turns out that there are some differences between the results obtained with the two models. Mainly, we notice that particle acceleration up to $\gamma_{\rm max}=21.6^{+39.0}_{-7.9}$ is sufficient to reproduce the observed high-energy tail. Moreover, a truncated electron distribution allows to improve the quality of the best fit to $\chi^2/74 = 1.00$. The total opacity of the plasma is found equal to the value obtained with a non-truncated distribution, but the pair yield is reduced due to the lack of very energetic particles. With respect to the ECM1, the other parameters are only marginally affected and remain consistent within the $90$ per cent confidence errors. 
\begin{table*}[tp]\footnotesize
\begin{minipage}{\textwidth}
\centering
\begin{tabular}{cccccccccccc}
\hline\hline
\T \B \multirow{2}*{Model} & \multirow{2}*{Spec} & \multirow{2}*{$\Gamma_{\rm{inj}}$} & \multirow{2}*{$\gamma_{\rm max}$} & \multirow{2}*{$l_{\rm nth}/l_{\rm s}$} & \multirow{2}*{$l_{B}$} & \multirow{2}*{$\tau_{\rm ion}$} & \multirow{2}*{$\tau_{\rm T}$} & $kT_{\rm e}$ & \multirow{2}*{$\Omega/2\pi$} & \multirow{2}*{$\xi$} & \multirow{2}*{$\chi^2_{\nu}$(d.o.f.)} \\
\B & & & & & & & & (keV) & & &
\\
\hline\hline
\T \multirow{2}*{ECM1} & cutoff & $3.95^{+1.55}_{-0.70}$ & $1000^{\star}$ & $4.35{\pm0.07}$ & --- & $2.98{\pm0.07}$ & $2.99{\pm0.07}$ & $20.1$ & $0.26^{+0.03}_{-0.01}$ & $710{\pm120}$ & $0.84 (68)$  \\ 
\T \B & excess & $2.55^{+0.35}_{-0.15}$ & $1000^{\star}$ & $5.06^{+0.10}_{-0.08}$ & --- & $3.31^{+0.07}_{-0.15}$ & $3.87^{+0.07}_{-0.15}$ & $14.3$ & $0.25^{+0.02}_{-0.03}$ & $600{\pm240}$ & $1.10 (74)$ 
\\
\T \multirow{2}*{ECM2} & cutoff & $2.5^{\star}$ & $4.0^{+1.3}_{-0.3}$ & $4.25^{+0.08}_{-0.10}$ & --- & $3.06^{+0.09}_{-0.06}$ & $3.07^{+0.09}_{-0.06}$ & $19.1$ & $0.23^{+0.03}_{-0.02}$ & $850^{+140}_{-200}$ & $0.85 (68)$  \\ 
\T \B & excess & $2.5^{\star}$ & $21.6^{+39}_{-7.9}$ & $4.94^{+0.11}_{-0.10}$ & --- & $3.69^{+0.04}_{-0.13}$ & $3.87^{+0.04}_{-0.13}$ & $14.5$ & $0.21^{+0.03}_{-0.02}$ & $1240^{+290}_{-240}$ & $1.00 (74)$ 
\\  \hline
\T \multirow{2}*{ICM1} & cutoff & $3.47_{-0.31}^{+0.66}$ & $1000^{\star}$ & --- & $740^{+200}_{-140}$ & $2.26{\pm0.18}$ & $2.26{\pm0.18}$ & $34.2$ & $0.43{\pm0.05}$ & $300^{+120}_{-100}$ & $0.96 (68)$  \\ 
\T \B & excess & $2.50^{+0.32}_{-0.08}$ & $1000^{\star}$ & --- & $25.2{\pm3.6}$ & $2.14^{+0.14}_{-0.17}$ & $2.45^{+0.14}_{-0.17}$ & $25.7$ & $0.44^{+0.06}_{-0.03}$ & $260^{+110}_{-80}$ &  $1.06 (74)$ 
\\ 
\T \multirow{2}*{ICM2} & cutoff & $2.5^{\star}$ & $15.2^{+5.5}_{-1.3}$ & --- & $283^{+43}_{-36}$ & $2.30^{+0.17}_{-0.16}$ & $2.30^{+0.17}_{-0.16}$ & $33.4$ & $0.43^{+0.04}_{-0.03}$ & $300^{+120}_{-100}$ & $0.96 (68)$  \\ 
\T \B & excess & $2.5^{\star}$ & $190^{+110}_{-105}$ & --- & $27.3^{+4.0}_{-3.4}$ & $2.49{\pm0.18}$ & $2.73{\pm0.18}$ & $23.9$ & $0.39^{+0.04}_{-0.04}$ & $310^{+150}_{-130}$ & $0.92 (74)$
\\
\hline
\label{tab2}
\end{tabular}
\caption{Best fit parameters of the joint SPI and PCA spectra for the non-thermal models. Fixed parameters are indicated by a star ($^{\star}$) next to their value. The temperature of the thermal component is estimated at the equilibrium using Equation 2.8 of \citet{coppi1992} and is hence not a fit parameter.}
\end{minipage}
\end{table*}

\subsection{Non-thermal acceleration with internal Comptonization}

Finally, we consider models in which the observed $X/\gamma$-ray spectra are produced in a magnetized plasma.
To emphasize the effects of the magnetic field, we study the case where all the seed photons are internally generated from synchrotron emission ($l_{\rm s}=0$, $l_B\neq0$). 
As in the previously analyzed non-magnetic models, we adopt the same different configurations to phenomenologically describe the acceleration mechanism.
For convenience, we call these models the ICM1 and the ICM2 (for Internal Comptonization Models), respectively.
To calculate the model spectra, we used the new versatile Comptonization code \textsc{belm} \citep*{bmm08}. One of the main differences with \textsc{eqpair} is that  \textsc{belm} accurately accounts for self-absorbed cyclo-synchrotron radiation from the sub-relativistic to the ultra-relativistic regime. We compared the spectra obtained by both codes for the best fit parameters of the non-magnetized models. The relative differences are smaller than 3 per cent at all energy. The fitting results are summarized in Table 3.
\begin{figure*}
\begin{minipage}{\textwidth}
\begin{tabular}{cc}
  \includegraphics[width=8.55cm]{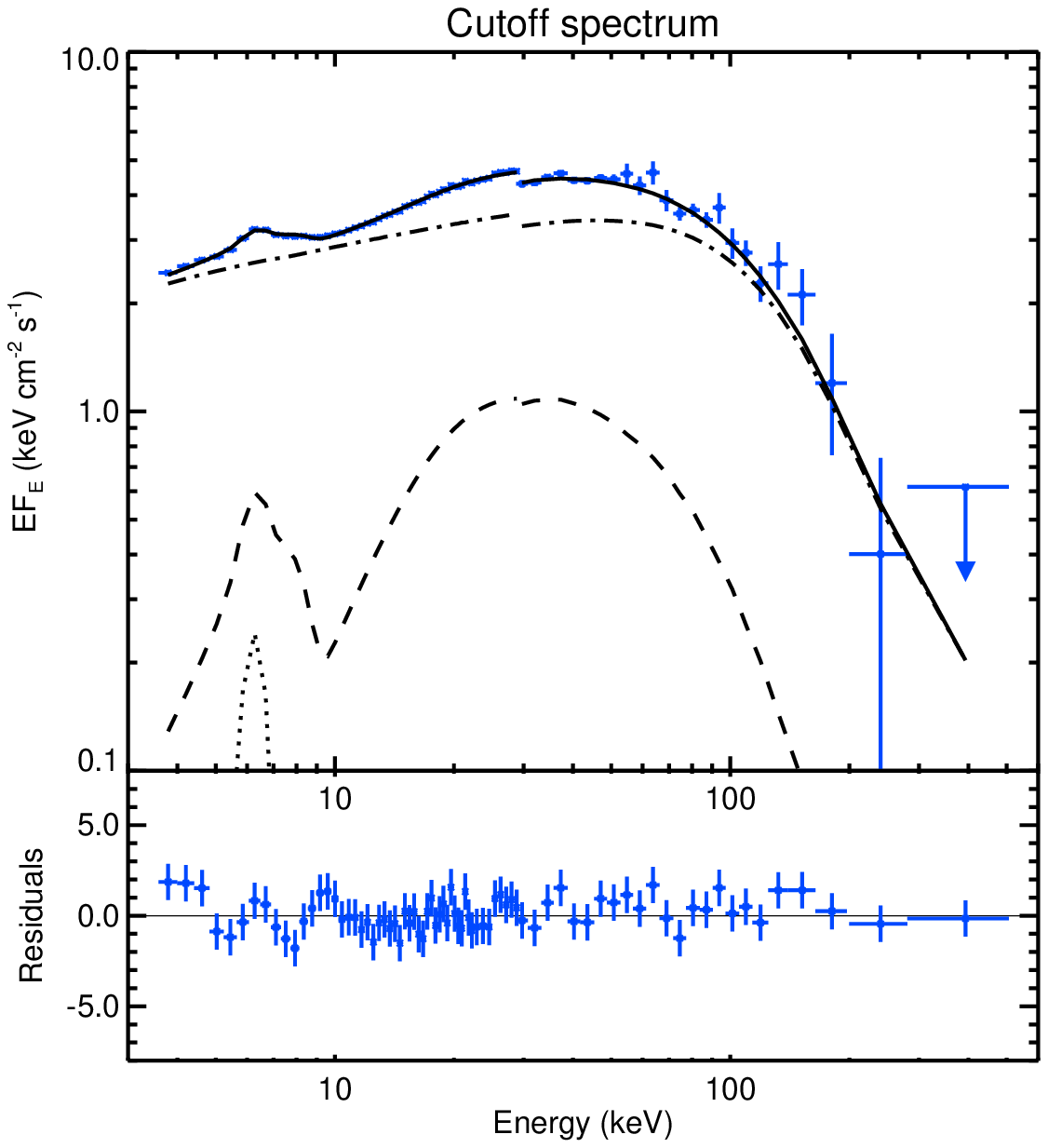} &
  \includegraphics[width=8.55cm]{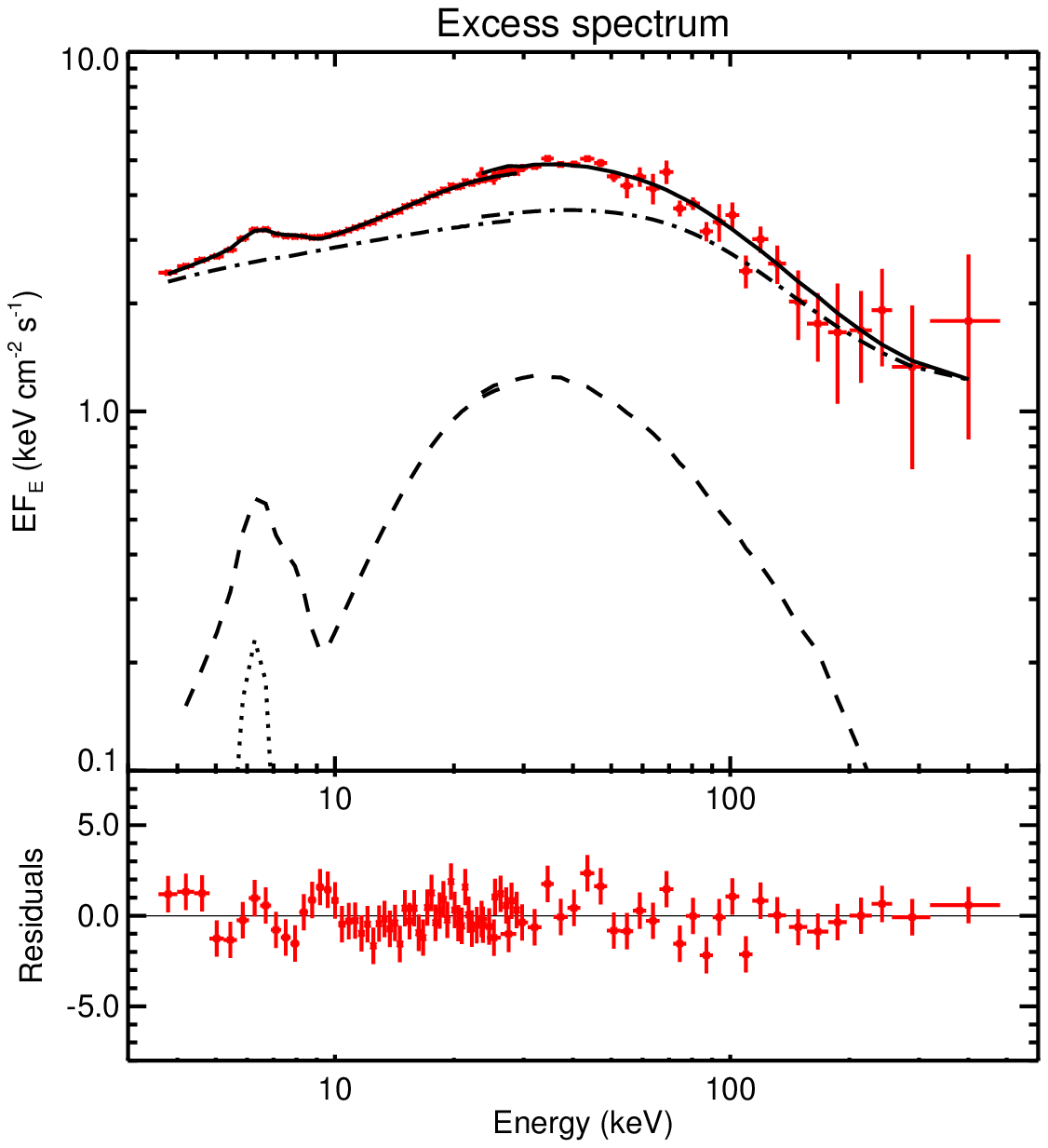}
\end{tabular}
  \centering
  \caption{Joint SPI and PCA spectra fitted with the ICM1. The solid line is the total model, the dot-dashed line represents the SSC component while the dashed and dotted lines show the reflection component and the iron line emission, respectively. The error bars are given at the $1\sigma$ level.}\label{ren_ginj}
\end{minipage}
\end{figure*}

\subsubsection{The injection index}

First, we employ the commonly used configuration, i.e. we fix $(\gamma_{\rm min},\gamma_{\rm max})=(1.3,1000)$.
A qualitative analysis shows that the model spectrum below $30$\,keV is rather insensitive to the individual values of $l_{\rm nth}$, $l_B$ and $\Gamma_{\rm{inj}}$, but strongly depends on a combination of all three parameters (see also MB09). The situation is similar as in the non-magnetic case (cf. section 4.2), although the dependence is slightly more complicated. On the other hand, the high-energy spectrum ($>$$100$\,keV) is mostly determined by the injection index $\Gamma_{\rm{inj}}$. As a consequence, we use Eq. \ref{comp} to fix $l_{\rm nth}=100$, while the broad band spectra allow to disentangle the degeneracy between $l_{B}$ and $\Gamma_{\rm{inj}}$.

Due to the detailed treatment of the microphysics in the \textsc{belm} code, real-time fits in \textsc{xspec} are very time consuming. In order to make the fitting process more efficient, we tabulated the model spectra\footnote{We used the \textsc{wftbmd} routine (publicly available at the heasarc website) to create the appropriate FITS file required by the \textsc{atable} model in \textsc{xspec}}. The resulting table file has three dimensions (corresponding to the parameters $l_{B}$, $\Gamma_{\rm{inj}}$ and $\tau_{\rm ion}$) and the fits are performed through interpolation between the tabulated spectra. 

In order to account for Compton reflection from a cold disc, the \textsc{belm} model is convolved with an angle-dependent reflection routine based on the \textsc{pexriv} model by \citet{magdziarz1995}, but taking into account the distortions due to general relativistic effects.
The free parameters of the ICM1 are hence $l_{B}$, $\Gamma_{\rm{inj}}$, $\tau_{\rm ion}$, $\Omega/2\pi$, $\xi$, the iron line energy and the normalization factor between PCA and SPI.

For the \textit{cutoff} spectrum, the ICM1 provides a good fit to the data ($\chi^2/68 = 0.96$). In comparison with the ECM1, the high-energy rollover is better reproduced.
However, as can be seen from the residuals in Figure \ref{ren_ginj} (left), the fit is slightly less accurate in the iron line region around $6.4$\,keV. There are also some positive residuals below $5$\,keV, possibly hinting the need for a soft disc component. Since we focus on the spectral behavior at high energies, we did not investigate these issues any futher.
The best fit is obtained with an electron spectral index of $\Gamma_{\rm inj}=3.47_{-0.31}^{+0.66}$.
The magnetic compactness is fitted at $l_B=740^{+200}_{-140}$, which corresponds to a magnetic field strength of $B=2.0^{+1.0}_{-0.9}\times 10^7$ G.
The plasma is found to be of moderate optical thickness, with fitted $\tau_{\rm ion}=2.26 \pm 0.18$.  Due to the fast synchrotron cooling of the small number of high-energy particles, the produced pair yield is negligible. As with the ECM1, the data strongly require ionized Compton reflection, with a fitted amplitude and ionization factor of $\Omega/2\pi=0.43\pm0.05$ and $\xi=300^{+120}_{-100}$, respectively.

For the \textit{excess} spectrum, the ICM1 allows again a good description of the data (cf Figure \ref{ren_ginj} right). The best fit yields $\chi^2/74 = 1.06$ and the high-energy data constrain the injection index to
$\Gamma_{\rm inj}=2.50^{+0.32}_{-0.08}$. This result is roughly equal to the value obtained with the ECM1. However, at equal injected electron distribution, the high-energy tail of the ICM1 spectrum is slightly steeper. Indeed, contrary to a non-magnetized model, the Compton losses compete with the synchrotron losses and a significant fraction of the energy radiated by the non-thermal leptons is emitted in the optical/UV range rather than in hard X-rays. Accounting for the the high-energy tail therefore requires more high-energy particles, i.e. a smaller $\Gamma_{\rm inj}$. We find a magnetic compactness of $l_B=25.2\pm3.6$, which given our assumptions translates to $B=3.76\pm0.14\times 10^6$ G. This means that in the framework of this model, the magnetic field strength would drop by a factor of $5.5$ during a change from the \textit{cutoff} to the \textit{excess} spectrum. 

The Thomson optical depth of ionization electrons is fitted at $\tau_{\rm ion}=2.14^{+0.14}_{-0.17}$. Contrary to the results obtained with the ECM1, the $\tau_{\rm ion}$ values for both spectra are compatible within the $90$ per cent confidence errors. The reflection routine yields $\Omega/2\pi=0.44^{+0.06}_{-0.03}$ and $\xi=260^{+110}_{-80}$, which means that these characteristics did not change either.
\subsubsection{The maximum particle energy}

In the ICM2, we keep $l_{\rm nth}=100$ and $\gamma_{\rm min}=1.3$, but allow now for variations of $\gamma_{\rm max}$. Instead, we set the injection index to $\Gamma_{\rm inj}=2.5$ and generate a new fitting table which has again three dimensions, corresponding to the free parameters  $l_{B}$, $\gamma_{\rm max}$ and $\tau_{\rm ion}$.

\begin{figure*}
\begin{minipage}{\textwidth}
\begin{tabular}{cc}
  \includegraphics[width=8.55cm]{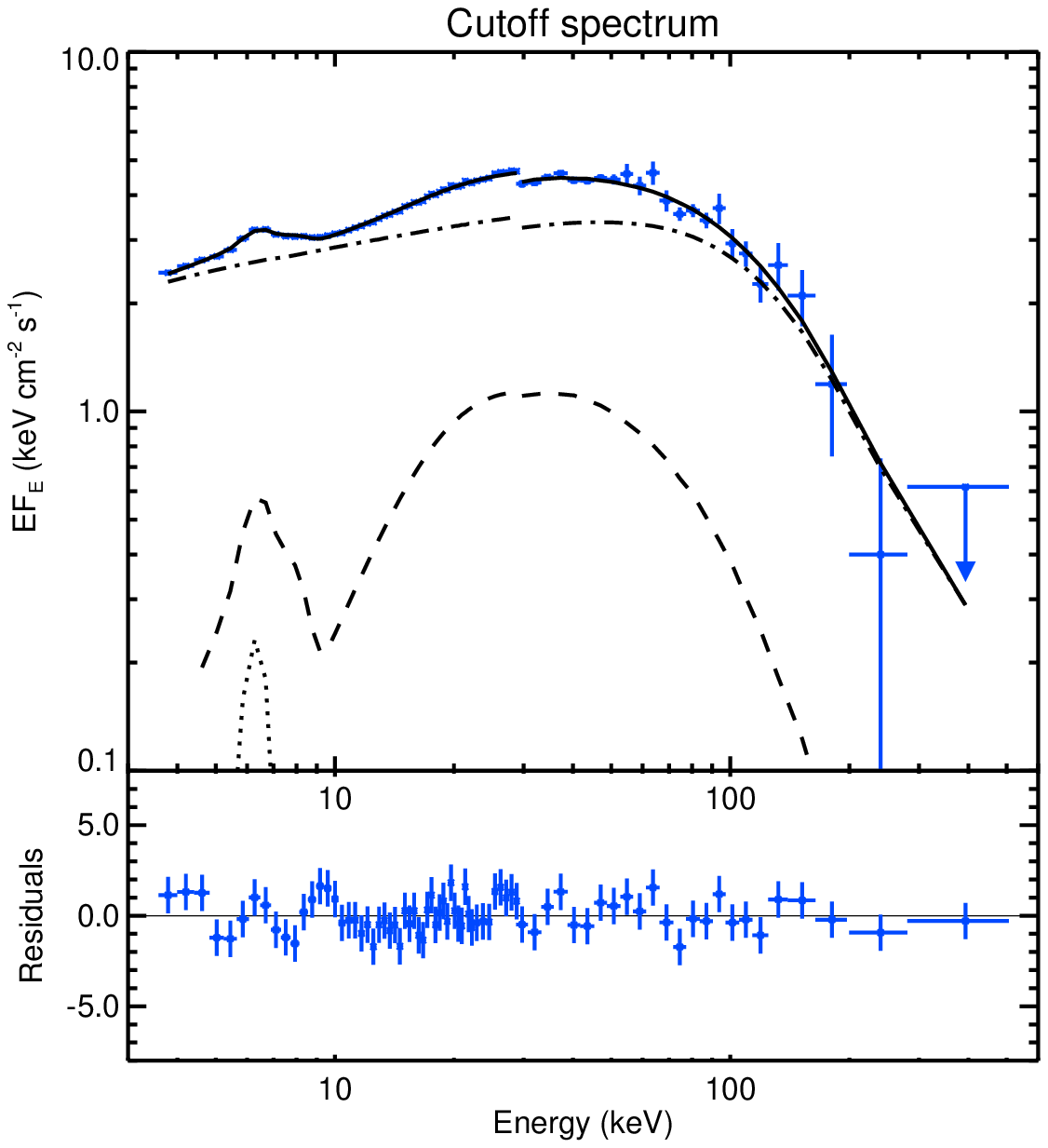} &
  \includegraphics[width=8.55cm]{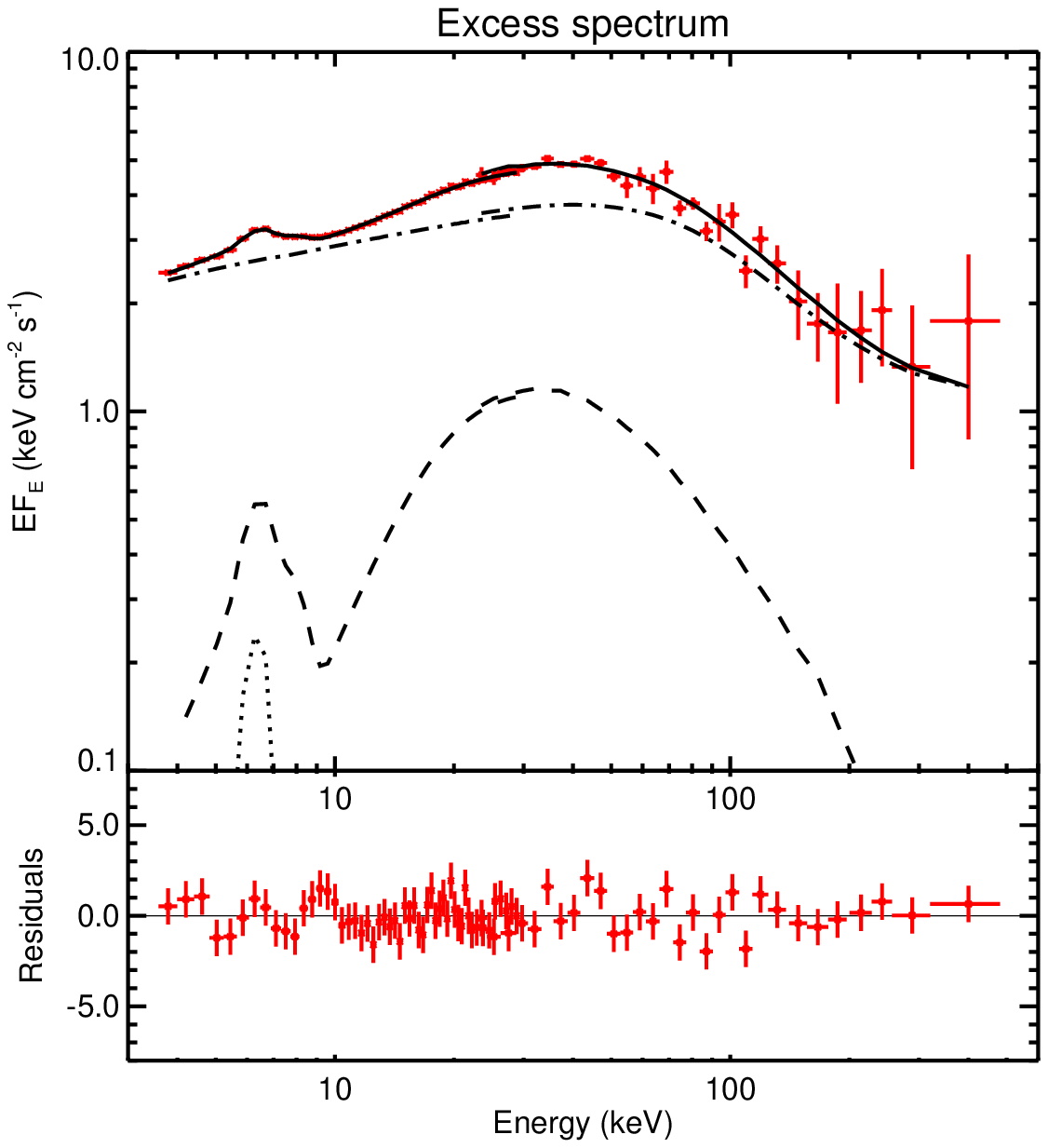}
\end{tabular}
  \centering
  \caption{Joint SPI and PCA spectra fitted with the ICM2. See Figure \ref{ren_ginj} for the description of the different model components. The error bars are given at the $1\sigma$ level.}\label{ren_gmax}
\end{minipage}
\end{figure*}

For the \textit{cutoff} spectrum, we obtain a good fit to the data ($\chi^2/68 = 0.96$; shown in Figure \ref{ren_gmax} left), qualitatively equivalent to the best fit obtained with the ICM1. Again, this shows that the effects of a very soft injection slope
can be mimicked by a much harder injected distribution, but truncated at a certain particle energy. We find $\gamma_{\rm max}=15.2^{+5.5}_{-1.3}$, which is significantly higher than the fitted maximum particle energy in the ECM2. The inferred magnetic compactness is reduced with respect to the ICM1, that is to say $l_B=280^{+43}_{-36}$, which for a medium of typical size $R=30$\,R$_{\rm G}$ corresponds to $B=1.25^{+0.5}_{-0.4}\times 10^7$\,G. The other parameters are not much affected, namely we find $\tau_{\rm ion}\simeq\tau_{\rm T}=2.30^{+0.16}_{-0.16}$ and from the reflected component we infer $\Omega/2\pi=0.43^{+0.04}_{-0.03}$ and $\xi=300^{+120}_{-100}$.

For the \textit{excess} spectrum, we note that the injection index is fixed to the best fit value obtained with the ICM1. However, allowing for $\gamma_{\rm max}$ to vary (i.e allowing for a truncated electron distribution), the fit may be improved considerably. For $\gamma_{\rm max}=190^{+110}_{-105}$, we obtain the best description of the \textit{excess} spectrum, with a reduced $\chi^2$ of $\chi^2/74 = 0.92$ (cf. Figure \ref{ren_gmax} right). Compared to the non-magnetized model, the maximum electron energy is again found to be much higher. This is expected since in the magnetic models, the seed photons have a lower average energy than in the non-magnetized models. Indeed, in the ICMs, the high-energy photons are produced from single Compton scattering off the synchrotron emission ($E_{\rm s} \sim 0.01$\,keV) while in ECMs the seed photons originate from the disc emission ($E_{\rm s} \sim 1$\,keV). Therefore, in order to upscatter the seed photons to $200$\,keV, the electrons must have averaged Lorentz factors of $\gamma\sim12$ in the ECMs and $\gamma\sim 120$ in the ICMs.

The magnetic compactness remains equal to the value inferred with the ICM1, namely $l_B=27.3^{+4.0}_{-3.4}$. A transition from the cutoff to the excess spectrum thus requires a factor $3.2$ decrease of the magnetic field strength. The opacity from ionization electrons yields $\tau_{\rm ion}=2.49\pm0.18$, while the total optical depth is found to be $\tau_{\rm T}=2.73\pm0.18$. This shows again that the ionization opacity does not change between the two spectra. Finally, the fitted reflection amplitude and ionization parameter are found to be consistent for both spectra regardless of the acceleration model (cf. Table 3). We thus conclude that in the framework of a strongly magnetized medium, the constraints to these parameters are very robust.

\section{Discussion}

The SPI observations showed that during the bright hard state of the 2007 outburst, the highest energy emission ($>$$150$\,keV) of GX 339--4 was variable. While the spectral shape at lower energies (4--150\,keV) remained more or less constant, we detected the significant appearance/disappearance of a high-energy tail. The strength of this hard tail, varying on a time scale of less than $7$\,hours, is found to be positively correlated with the total X-ray luminosity of the source and enables interesting constraints to the physical processes which could be responsible for the high-energy emission.

\subsection{The pure thermal model}

The clear detection of a cutoff energy in the \textit{cutoff} spectrum indicates that the Comptonizing electron distribution is quasi Maxwellian. 
Thus, it is not surprising that the spectrum can be explained by assuming only thermal heating of the plasma. As suggested by the advection dominated accretion flow models, this could be achieved through Coulomb interactions with a thermal distribution of hot protons. The temperature of the protons can be estimated from the thermal compactness, the electron temperature and the optical depth of the plasma (cf. formula (4) in MB09). For $l_{\rm th}=100$, we infer a proton temperature of the order of $1$\,MeV. This is of the same order of magnitude than the proton temperature estimated in MB09 for the canonical hard state of Cygnus X-1, which in comparison to the hard state of GX 339--4 analyzed here shows a hotter electron plasma ($kT_{\rm e} \sim 85$\,keV) but a lower compactness ($l_{\rm th} \sim 5$). As mentioned in MB09, proton temperatures of the order of $1$\,MeV are significantly lower than what is expected in typical two-temperature accretion flows, namely $kT_{\rm i} > 10$\,MeV.

In any case, pure thermal heating is not enough since it is not able to explain the appearance of the observed high-energy tail.
We conclude that either the hard excess is independent from the thermal component, in which case its origin is located outside the innermost regions, or that both components are linked, in which case at least some level of non-thermal heating is required.

On the other hand, both broad band spectra can be successfully explained by models involving \textit{only} non-thermal electron acceleration. This suggests that the Comptonizing medium in hard states could be powered by the same non-thermal mechanisms that are believed to power the accretion disc corona in soft states. Depending on the nature of the plasma (magnetized or not) and the origin of the seed photons, these issues are discussed in the next paragraphs.

\subsection{The non-magnetic case}

In the framework of a non-magnetized model, the variability of the high-energy spectrum can be explained by changes in the properties of the involved acceleration processes.
Namely, the fits suggest a small variation of the total power supplied to the plasma along with a significant variation of either the spectral index (in the ECM1) or the cutoff energy (in the ECM2) of the non-thermal electron distribution.

In the ECM1, a change from the \textit{cutoff} to the \textit{excess} spectrum requires that the spectral index typically drops from $\Gamma_{\rm inj}=4.0$ to $2.5$. This implies that the average energy of the accelerated particles $<$$E_{\rm inj}$$>$ rises from $1.0$ to $1.8$\,MeV. 
In the ECM2, at constant spectral index, the best fits show that relatively low maximum Lorentz factors are sufficient to reproduce the data. In this case, the appearance of the high-energy excess requires an increase of $\gamma_{\rm max}$ from $4$ to $22$, which implies that the average energy of the accelerated electrons rises accordingly, from $1.0$ to $1.5$\,MeV. 
In both cases, such variations can be explained by the possible non-stationarity of the inherent acceleration mechanisms. For instance, in the framework of shock acceleration, the properties of the accelerated particles depend on the shock strength \citep{webb1984,spitkovsky2008}. Acceleration by reconnection depends on several physical parameters such as the local geometry of the reconnection zone \citep{zenitani2007} and the number of reconnection sites (if the particles are accelerated stochastically by successive acceleration events in different sites \citep{anastasiadis1997,dauphin2007}). All these properties may undergo variations with time and could hence explain the observed variability.

\begin{figure*}
\begin{minipage}{\textwidth}
\begin{tabular}{cc}
  \includegraphics[width=8.55cm]{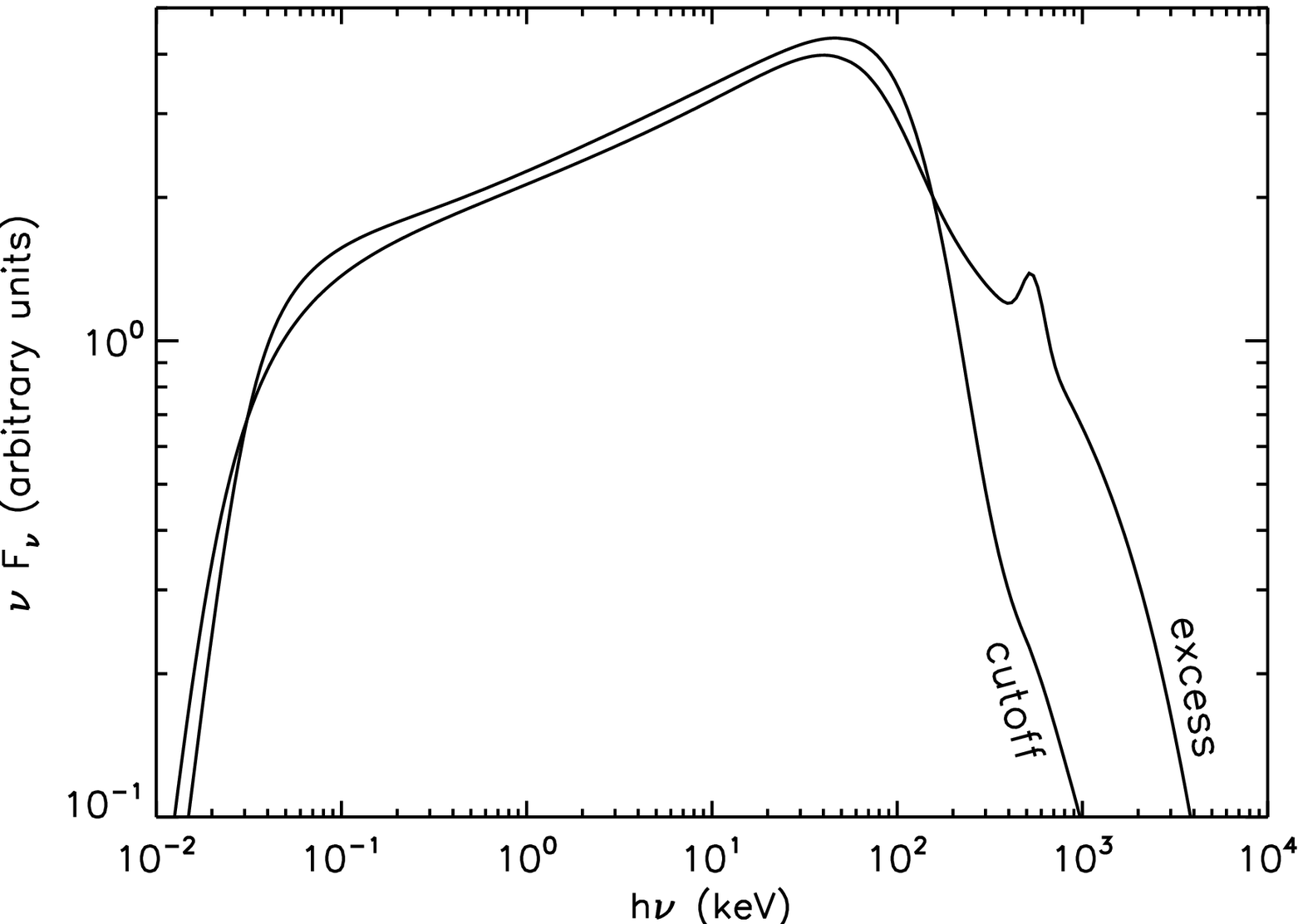} &
  \includegraphics[width=8.55cm]{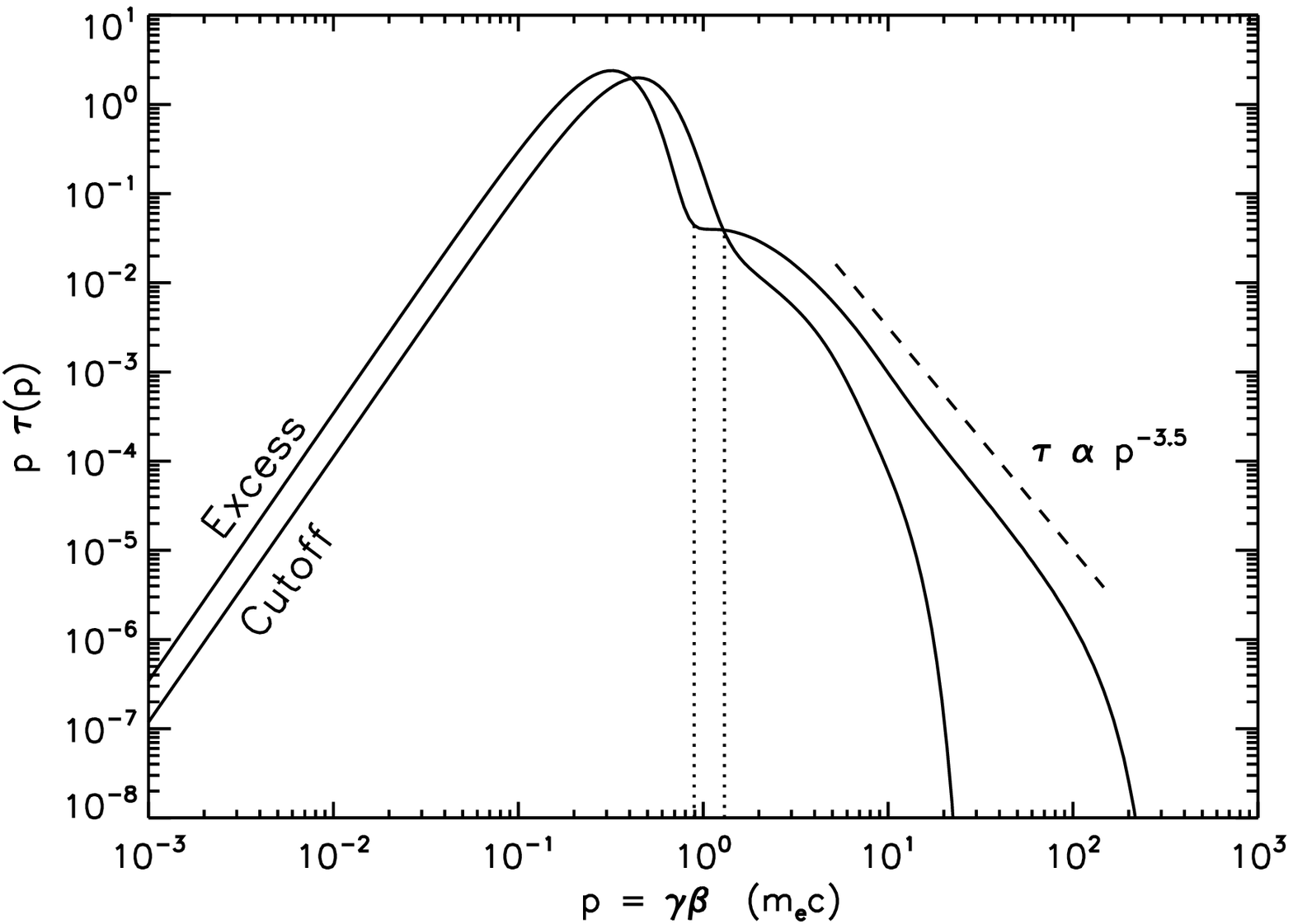}
\end{tabular}
  \centering
  \caption{Photon spectra (left panel) and particle distributions (right panel) in the \textit{cutoff} and \textit{excess} states, obtained with the \textsc{belm} code for the best fit parameters of the ICM2. The photon spectra include neither absorption nor reflection. $\tau(p)$ is the Thomson optical depth per unit momentum. For each particle distribution, the vertical dotted lines show the energy chosen to define the thermal and the non-thermal populations (cf. section 5.3).}\label{spec_part}
\end{minipage}
\end{figure*}

Our results are consistent with those of \citet{gierlinski2005}, who studied the energy dependent variability of non-magnetized Comptonization models in response to varying physical parameters. Although the variations of $\Gamma_{\rm inj}$ and $\gamma_{\rm max}$ could not explain the observed rms spectra of XTE J1650--500 and XTE J1550-564, they found that changes in these parameters produce strong variations in the X-ray spectrum above 50 keV, which is what is required here to reproduce the present data.

The spectral analysis suggests that a transition from the \textit{cutoff} to the \textit{excess} spectrum additionally requires a $15$ per cent increase of the total power supplied to the plasma. Considering a spherical medium of fixed radius and constant illumination from the accretion disc, this increase is consistent with the $14$ per cent difference in the observed 4--500\,keV luminosity. In addition, the total optical depth of the plasma is found to increase by about $30$ per cent, which is mainly due to the enhanced pair production occuring in the harder acceleration regime.

Motivated by size and luminosity estimates, we assumed a constant soft photon compactness of $l_{\rm s}=15$. As mentioned earlier, the above results are only weakly dependent on the exact value of $l_{\rm s}$. Very high- or very low values of the illumination compactness, however, turned out to be inconsistent with the observed spectra. Indeed, since the ratio $l_{\rm nth}/l_{\rm s}$ is robustly constrained, strong illumination requires an efficient acceleration mechanism (i.e. large $l_{\rm nth}$) to reproduce the spectral slope at lower energies ($<20$\,keV). Consequently, this generates very energetic radiation which produces large amounts of $e^-$$e^+$ pairs through photon-photon annihilation. This, in turn, reduces the equilibrium temperature of the plasma since more particles have to share the same amount of energy. Hence, above a certain soft photon compactness, the equilibrium temperature will be too low to be consistent with the observed thermal peak of the spectrum. 

Reciprocally, since the thermal part of the electron distribution is roughly determined by the balance between acceleration and Compton cooling, decreasing $l_{\rm s}$ at constant $l_{\rm nth}/l_{\rm s}$ has no significant effect on the lower energy part of the photon spectrum. However, a major fraction of the high-energy tail results from the Comptonization by mildly relativistic particles ($2<\gamma<10$). Below a certain soft photon compactness, the cooling of these mildly relativistic particles is no longer dominated by the Compton losses but by Coulomb interactions with the lower-energy thermal electrons. Thus, decreasing $l_{\rm s}$ at constant $l_{\rm nth}/l_{\rm s}$ provides a weaker acceleration rate while the cooling remains constant. This reduces the intensity of the high-energy tail up to the point where the model predictions are no longer consistent with the data. 

In conclusion, independently of any geometric argument, we obtain conservative bounds to the total compactness of the X-ray emitting plasma, i.e. $2<l<1500$. These limits are consistent with the estimates derived from geometric arguments, but unfortunately not very constraining. Anyhow, the robustness of the fitted compactness ratio $l_{\rm nth}/l_{\rm s} \simeq 4.5$ allows to conclude that the luminosity of the cold disc represents at most $\sim$\,$20$ per cent of the luminosity of the Comptonized component, possibly much less if the plasma is magnetized.

\subsection{The magnetic case}

Using the new code \textsc{belm} \citep*{bmm08}, we showed that the hard X-ray behavior of GX 339--4 in the bright hard state can be explained by assuming pure non-thermal electron acceleration and subsequent Comptonization of the self-consistently produced synchrotron photons. The model requires no incident radiation from the accretion disc and assumes constant power injection into the magnetized plasma. As in the non-magnetic case, the spectral variability can be mimicked by two different configurations of the acceleration model, involving either a variable power law slope (in the ICM1) or a variable maximum energy (in the ICM2) of the injected electron distribution.

In principle, these models allow to estimate the averaged magnetic field strength of the plasma. However, since the fits provide precise constraints only for the ratio $l_{B}/l$, the uncertainties on the total compactness (cf. equation (\ref{comp})) are projected to the estimate of $l_B$. To discuss our results, we express $l_B$ as a fraction of the magnetic compactness at equipartition with the radiation field $l_{B_{\rm R}}$. As we have the approximate dependence $l_{B_{\rm R}} \propto l\times (1+\tau_{\rm T}/3)$ (cf. equation (8) in MB09), the ratio $l_{B}/l_{B_{\rm R}}$ does not depend on the uncertainties regarding the source size and distance and is therefore a good indicator to quantify the importance of the magnetic processes.
In addition, $l_{\rm s}=0$ was assumed to study the physics of a strongly magnetized medium, but it can not be excluded that both the synchrotron flux \textit{and} a soft disc component contribute to the cooling of the non-thermal particles. If the medium is additionally illuminated by cold disc photons, less synchrotron cooling will be required to reproduce the slope of the lower energy spectrum, implying that the fitted values of the magnetic compactness are in fact conservative upper limits.
In the ICM1, we infer $l_{B}/l_{B_{\rm R}}\leq18$ from the \textit{cutoff} and $l_{B}/l_{B_{\rm R}}\leq0.58$ from the \textit{excess} spectrum. In the ICM2, the fitted magnetic compactness for the \textit{cutoff} spectrum is lower, namely we find $l_{B}/l_{B_{\rm R}}\leq6.0$ and $l_{B}/l_{B_{\rm R}}\leq0.60$, respectively.

As mentioned above, the magnetic models do not require any disc blackbody photons to produce the observed 4--500\,keV spectra. 
If the accretion disc extends down close to the black hole (as requested by the accretion disc corona models), the paucity of soft disc photons can be explained if the Comptonizing coronal material is outflowing at a mildly relativistic speed \citep{beloborodov1999,malzac2001}. Using the formulae (5) and (7) of \citet{beloborodov1999}, we estimate that bulk velocities of at least $0.6\,c$ are required for the cooling of the corona being dominated by synchrotron self-Compton. Comptonization off a dynamical corona may then blueshift the emerging spectrum, but these corrections remain moderate at $\sim$$0.6\,c$ and have not been included in the spectral fits.
On the other hand, the disc may as well be truncated (as requested by the hot flow models), since the data do not explicitly require relativistic smearing of the reflection features. Thus, as long as the particle acceleration is essentially non-thermal, the ICMs may apply to both geometries.

If the electron cooling is dominated by synchrotron self-Compton ($l_{\rm s}/l_{\rm nth}\ll 1/5$ from the non-magnetic models), the fits suggest that the magnetic field strength is roughly in equipartition with the radiation field. From a qualitative fit of the canonical hard state spectrum of Cygnus X-1, MB09 constrained the magnetic energy density to be strictly below equipartition ($l_{B}/l_{B_{\rm R}}<0.3$). As a consequence, even if our results are consistent with the results obtained for Cygnus X-1, the magnetic field could play a more important role in the physics of GX 339--4, at least in the cutoff state.

In both magnetic models, a transition between the two spectra can be explained by the variations of only two parameters. In any case, the magnetic compactness $l_B$ needs to be variable, along with either the injection slope (in the ICM1) or the maximum energy of the accelerated particles (in the ICM2). All other fit parameters are found to remain constant within the $90$ per cent confidence errors. To reproduce the \textit{cutoff} to \textit{excess} transition with the ICM1, the magnetic field has to decrease by a factor of $5.4$ and the injection slope drops from $\Gamma_{\rm inj}=3.5$ to $2.5$, while in the ICM2, the magnetic field decreases by a factor of $3.2$ and the maximum energy increases from $\gamma_{\rm max}=15.2$ to $187$.

Regardless of the precise acceleration process involved in the accretion flow, the inferred change in the magnetic field strength is expected to have an impact on the cutoff energy of the accelerated electron distribution. Indeed, the maximum energy of the particles is achieved when the energy losses become larger than the gains. In the magnetized models investigated in this paper, the energy losses of the most energetic particles are dominated by synchrotron cooling (Compton and Coulomb cooling are significantly smaller), which obviously depends on the magnetic field strength. In the framework of diffusive shock acceleration for instance, it has been shown that when synchrotron losses are included, the maximum Lorentz factor of the accelerated relativistic electrons satisfies $\gamma_{\rm max} \propto 1/(K B^2)$, where $K$ is the diffusion coefficient (see e.g. \citealt{webb1984} or \citealt{marcowith1999}). If $K$ is constant, this predicts a maximum energy of $\gamma_{\rm max} \propto 1/B^2$. However, depending on the assumptions made to describe the acceleration mechanism, the diffusion coefficient can depend both on the particle energy and the magnetic field strength: $K(\gamma,B)$. In the frequently used Bohm limit, it is assumed that $K\propto \gamma /B$, implying that the maximum Lorentz factor of the accelerated particles is expected to follow $\gamma_{\rm max}\propto B^{-1/2}$.
Although our results are not consistent with the Bohm predictions, the overall behaviour remains that the cutoff energy decreases with an increasing magnetic field strength. Moreover, other acceleration mechanisms, such as magnetic reconnection for instance, may give different predictions. 

As mentioned in the previous section, the injection index is not universal and may undergo variations in response to changing physical conditions in the acceleration region. However, while mere synchrotron cooling implies that $\gamma_{\rm max}$ and $B$ are necessarily anti-correlated, the physical connection between the variations of $\Gamma_{\rm inj}$ and the magnetic field strength remains less obvious. Hence, in the context of a magnetized plasma, the model with variable $\gamma_{\rm max}$ is slightly favored.
Anyhow, the variability of the acceleration mechanism was modeled by the variations of only one parameter (either the injection index or the cutoff energy), although it cannot be excluded that both parameters undergo simultaneous variations. We emphasize however that such models are more complicated for they imply more varying parameters and would not significantly improve the quality of the fits.

\begin{figure*}
\begin{minipage}{\textwidth}
  \includegraphics[width=17cm]{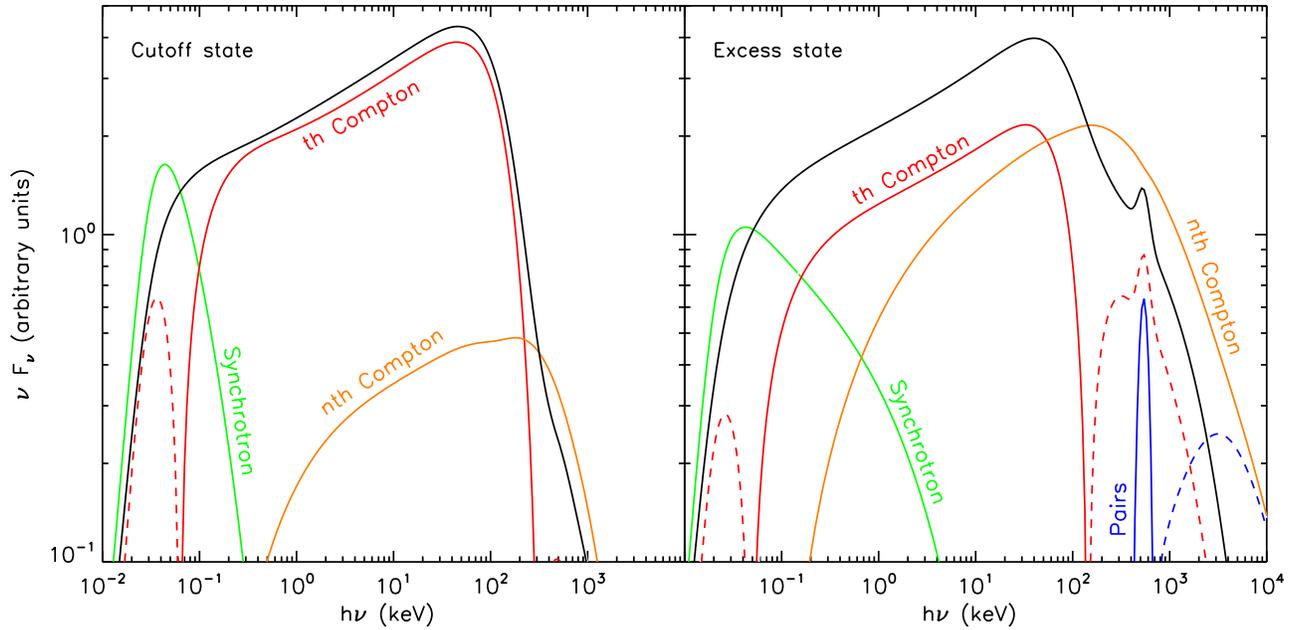} 
  \centering
  \caption{Photon spectra obtained with the \textsc{belm} code for the best fit parameters of the ICM2 in the \textit{cutoff} state (left panel) and the \textit{excess} state (right panel) with contributions from the various radiation processes: self-absorbed synchrotron emission (green lines), Comptonization off the thermal and non-thermal populations (red and orange lines, respectively) and pair production/annihilation (blue lines). Solid and dashed lines show positive and negative contributions to the spectra, respectively.}
\label{rad_mec}
\end{minipage}
\end{figure*}

In the context of the ICM2, we investigated the impact of the parameter variations on the electron distribution and the involved radiation mechanisms. Although the lower energy part of the model photon spectra is similar in shape, the underlying electron distributions are quite different (cf. Figure \ref{spec_part}). Indeed, the inferred variations of $B$ and $\gamma_{\rm max}$ not only change the non-thermal part of the distribution but also the temperature of the thermalized component. Figure \ref{rad_mec} compares the contributions from the various radiation mechanisms to the model photon spectra, showing separately the Comptonization off thermal and non-thermal particles\footnote{The low energy part of the particle distributions were fitted with a Maxwellian and particles of energy one order of magnitude larger than the inferred temperature were considered to be non-thermal.}. As expected, the parameter variations strongly increase the ratio of the non-thermal to the thermal component, resulting in the production of the high-energy tail.

In conclusion, the ICM2 provides a framework for a simple, physically motivated interpretation of the data, showing that the spectral variability could be triggered by the variation of only one single parameter, namely the magnetic field strength. Assuming a spherical medium of radius $R=30$\,R$_{\rm G}$, a black hole mass of $M=13$\,M$_{\sun}$ and a distance of $d=8$\,kpc, we infer a factor $3$ variation of the magnetic field strength, between $B\leq 3.9^{+1.5}_{-1.3}\times 10^6$\,G and $B\leq 1.25^{+0.50}_{-0.45}\times 10^7$\,G.
If the fitted average spectra provide a good estimate of the individual spectra from the single science windows, the time scale of this evolution is at most of the order of hours. Using the standard $\alpha$-prescription \citep{ss73}, the viscous time scale of the accretion disc at a radius $R$ is given by $t_{\rm visc} = \alpha^{-1} \left(H/R\right)^{-2} t_{\rm K}$, where $H/R$ is the aspect ratio of the disc and $t_{\rm K}$ the Keplerian period. Using the lower limit $\alpha \ge 0.01$ (quiescent disc) and $H/R \simeq 0.1$ (thin disc), we obtain $t_{\rm visc} < 10$\,min for the typical source size $R=30$\,R$_{\rm G}$. The viscous timescale of ADAF-like models is much shorter.
Therefore, even if the geometry of the X-ray emitting region and its dynamical evolution remain uncertain, global changes on time scales of hours are not unrealistic.

\section{Summary and Conclusion}

We presented an analysis of the high-energy emission of GX 339--4 in a luminous hard state. With respect to the standard cutoff shape of the hard state spectrum, the 25 -- 500\,keV \textit{INTEGRAL}/SPI data revealed the appearance of a variable high-energy excess. The intensity of this hard excess seems to be positively correlated with the total X-ray luminosity and the associated time scale is shorter than $7$\,hours. We explored the possible physical origins of this variability through an extensive analysis of two averaged spectra, one showing the typical cutoff shape and one showing this prominent high-energy excess.

We used simultaneous \textit{RXTE}/PCA data to extend the spectral coverage down to $4$\,keV and fitted the broad band spectra with a variety of physical Comptonization models. Models involving only thermal heating can be ruled out since they are not able to reproduce the high-energy tail. This feature thus confirms that in luminous hard states, the Comptonizing plasma of GX 339--4 contains a fraction of non-thermal particles. Models involving only non-thermal electron acceleration, on the other hand, showed that the thermal part of the spectrum can be produced by an initially non-thermal (power law) distribution, which rapidly thermalizes under the effects of synchrotron self-absorption and/or \textit{e-e} Coulomb collisions.

The relatively good signal to noise ratio of the high-energy channels ($>$$150$\,keV) allowed to derive meaningful constraints to the model parameters. Depending on the nature of the plasma (magnetized or not), the transition between the two averaged spectra requires the variations of at least two parameters. We found that a magnetized medium subject to non-thermal electron acceleration provides the framework for a simple and physically consistent interpretation of the data. Indeed, we showed that the spectral variability can be triggered by the \textit{only} variations of the magnetic field, implying a subsequent variation of the maximum energy of the accelerated particles. 

The quantitative constraints derived from this model yield a very conservative upper limit on the average magnetic field strength in the Comptonizing plasma
and suggest that in the bright hard state, the magnetic energy density could reach equipartition with the radiative energy density.

In conclusion, the presented results suggest that magnetic processes are likely to play a crucial role in the production of the high-energy emission of GX 339--4. In luminous hard states, the Comptonized emission could originate from a magnetized corona essentially powered though non-thermal particle acceleration, similarily to what is believed to happen in soft states. The hard X-ray emission in both spectral states may therefore be the consequence of a common physical phenomenon.

\acknowledgments
The SPI project has been completed under the responsibility
and leadership of the CNES. The authors are grateful to the ASI, CEA, DLR,
ESA, INTA, NASA, and OSTC for support. They also acknowledge financial support from CNRS, ANR and GDR  PCHE in France.
The authors thank the \textit{Swift}/\textit{BAT} and \textit{RXTE}/\textit{ASM} teams for publicly providing the monitor results. R.D. thanks M. del Santo and the IBIS team at INAF/IASF-Roma
for their kind assistance with the IBIS/ISGRI data analysis. We thank S. Motta for providing the reduced HEXTE data and C. Cabanac for valuable discussions. Last but not least we thank the anonymous referee whose comments substantially improved the quality of the paper.
\bibliography{biblist} 
\bibliographystyle{abbrvnat}

\end{document}